# How much feedback is required in MIMO Broadcast Channels?

Alireza Bayesteh, and Amir K. Khandani

Dept. of Electrical Engineering

University of Waterloo

Waterloo, ON, N2L 3G1

alireza, khandani@shannon2.uwaterloo.ca

## Abstract

In this paper, a downlink communication system, in which a Base Station (BS) equipped with $M$ antennas communicates with $N$ users each equipped with $K$ receive antennas ($K \leq M$), is considered. It is assumed that the receivers have perfect Channel State Information (CSI), while the BS only knows the partial CSI, provided by the receivers via feedback. The minimum amount of feedback required at the BS, to achieve the maximum sum-rate capacity in the asymptotic case of $N \to \infty$ is studied. First, the amount of feedback is defined as the average number of users who send information to the BS. For fixed SNR values, it is shown that with finite amount of feedback it is not possible to achieve the maximum sum-rate. Indeed, to reduce the gap between the achieved sum-rate and the optimum value to zero, a minimum feedback of $\ln \ln \ln N$ is asymptotically necessary. Next, the scenario in which the amount of feedback is defined as the average number of bits sent to the BS is considered, assuming different ranges of Signal to Noise Ratio (SNR). In the fixed and low SNR regimes, it is demonstrated that to achieve the maximum sum-rate, an infinite amount of feedback is required. Moreover, in order to reduce the gap to the optimum sum-rate to zero, in the fixed SNR regime, the minimum amount of feedback scales as $\Theta(\ln \ln \ln N)$, which is achievable by the Random Beam-Forming scheme proposed in [14]. In the high SNR regime, two cases are considered; in the case of $K < M$, it is proved that the minimum amount of feedback bits to reduce the gap between the achievable sum-rate and the maximum sum-rate to zero grows logaritmically with SNR, which is achievable by the "Generalized Random Beam-Forming" scheme, proposed in [18]. In the case of $K = M$, it is shown that by using the Random Beam-Forming scheme and the total amount of feedback not growing with SNR, the maximum sum-rate capacity is achieved.

I. INTRODUCTION

Multiple-Input Multiple-Output (MIMO) systems have proved their ability to achieve high bit rates in a scattering wireless network. In a point-to-point scenario, it has been shown that the capacity scales linearly with the minimum number of transmit and receive antennas, regardless of the availability of Channel State Information (CSI) at the transmitter [1] [2]. This linear increase is so-called *multiplexing gain*.

In a MIMO Broadcast Channel (MIMO-BC), a BS equipped with multiple antennas communicates with several multiple-antenna users. Recently, there has been a lot of interest in characterizing the capacity region of this channel [3], [4], [5], [6]. In these works, it has been shown that the sum-rate capacity of MIMO-BC grows linearly with the minimum number of transmit and receive antennas, provided that both transmitter and receiver sides have perfect CSI. Indeed, in a network with a large number of users, the BS can increase the throughput by selecting the best set of users to communicate with. This results in the so-called *multiuser diversity gain* [7], [8].

Unlike the point-to-point scenario, in MIMO-BC it is crucial for the transmitter to have CSI. It has been shown that MIMO-BC without CSI at the BS is degraded [9]. Moreover, for the case of single antenna users, multiplexing gain reduces to one, and multiuser diversity gain disappears [10] [11].

Due to the weak performance of having no CSI at the BS, some authors have considered MIMO-BC with partial CSI [10] [12] [13] [14] [15] [16] [17] [18]. In [12], the authors have proposed a user selection strategy in a single-antenna broadcast channel, which exploits the maximum sum-rate capacity with only one bit feedback per user. This idea has been generalized for MIMO-BC in [13], using the idea of antenna selection.

Reference [14] proposes a downlink transmission scheme based on random beam-forming, relying on partial CSI at the transmitter. In this scheme, the BS randomly constructs $M$ orthogonal beams and transmits data to the users with the maximum Signal to Interference plus Noise Ratio (SINR) for each beam. Therefore, only the value of maximum SINR, and the index of the beam for which the maximum SINR is achieved, are fed back to the BS for each user. This significantly reduces the amount of feedback. Reference [14] shows that when the number of users tends to infinity, the optimum sum-rate throughput can be achieved.

Reference [10] considers a downlink channel where a transmitter with $M$ antennas communicates with $M$ single-antenna receivers. It is assumed that receivers have perfect CSI, but the transmitter only has the quantized information regarding the channel instantiation. This reference shows that assuming Zero-Forcing Beam-Forming (ZFBF) precoding at the transmitter, the full multiplexing gain can be achieved with partial CSI, if the quality of the CSI is increased linearly with the SNR. This result is generalized in [15] to the case of multiple-antenna receivers, when the number of receive antennas is less than $M$. In [16], the authors consider a MIMO-BC when a transmitter with two antennas transmits data to two single-antenna receivers. They show that if the transmitter has the channel state with finite precision, the maximum achievable multiplexing gain is $\frac{2}{3}$ [1]. In fact, references [10], [15], and [16] study the performance degradation of MIMO-BC due to the imperfect CSI, at the high SNR regime. The size of the network (the number of users) is assumed to be fixed in these references.

In [17], we have considered a downlink scheme based on ZFBF and have proved that when the number of users, $N$, tends to infinity, the maximum sum-rate capacity is achievable with the amount of feedback scaling as $[\ln N]^M$. In [18], the authors have considered a MIMO-BC with large number of users at high SNR. They have shown that it is possible to achieve the maximum multiplexing gain with the amount of feedback per user decreasing with $N$. However, it is still required that the feedback load per user grows logaritmically with SNR. Two essential questions arise here: i) Is it possible to achieve the maximum sum-rate capacity with finite feedback in a large network ($N \to \infty$)? ii) If not, what is the minimum feedback rate (in terms of $N$ and SNR) in order to achieve the sum-rate capacity of the system?

In this paper, we aim to answer the above questions. First, we define the amount of feedback as the average number of users who send information to the BS. In the fixed and low SNR regimes, our results show that it is not possible to achieve the maximum sum-rate with a finite amount of feedback. Moreover, in the fixed SNR regime, in order to reduce the gap between the achieved sum-rate and the optimum value to zero, the amount of feedback must be greater than $\ln \ln \ln N$. In the second part, we define the amount of feedback as the number of information bits sent to the BS. In the fixed SNR regime, our analysis shows that the minimum amount of feedback, in order to reduce the gap to the optimum sum-rate to zero, scales as $\Theta(\ln \ln \ln N)$, which can be achieved using the Random Beam-Forming scheme proposed in [14]. However, the optimality of Random Beam-Forming only holds for the region $\ln P \nsim \Omega(\ln N)$. In the regime of $\ln P \sim \Omega(\ln N)$, we consider two cases. In the case of $K < M$, we prove that the minimum amount of feedback bits to reduce the gap between the achievable sum-rate and the maximum sum-rate to zero grows logaritmically with SNR, which is achievable by the "Generalized Random Beam-Forming" scheme, proposed in [18]. In the case of $K = M$, we show that by using the Random Beam-Forming scheme and the amount of feedback not growing with SNR the maximum sum-rate capacity is achievable.

---

[1]It is assumed that the transmitted signal and the channel coefficients are real.

In section II of this paper, we introduce the system model, while section III is devoted to the asymptotic analysis of the amount of feedback. Section IV concludes the paper.

Throughout this paper, the norm of the vectors and the Frobenius norm of the matrices are denoted by $\|.\|$. The Hermitian operation is denoted by $(.)^H$ and the determinant and the trace operations are denoted by $|.|$ and $\text{Tr}(.)$, respectively. $\mathbb{E}\{.\}$ represents the expectation, notation "ln" is used for the natural logarithm, and the rates are expressed in *nats*. RH(.) represents the right hand side of the equations. Indeed, for any functions $f(N)$ and $g(N)$, $f(N) = O(g(N))$ is equivalent to $\lim_{N \to \infty} \left|\frac{f(N)}{g(N)}\right| < \infty$, $f(N) = o(g(N))$ is equivalent to $\lim_{N \to \infty} \left|\frac{f(N)}{g(N)}\right| = 0$, $f(N) = \Omega(g(N))$ is equivalent to $\lim_{N \to \infty} \frac{f(N)}{g(N)} > 0$, $f(N) = \omega(g(N))$ is equivalent to $\lim_{N \to \infty} \frac{f(N)}{g(N)} = \infty$, $f(N) = \Theta(g(N))$ is equivalent to $\lim_{N \to \infty} \frac{f(N)}{g(N)} = c$, where $0 < c < \infty$, $f(N) \sim g(N)$ is equivalent to $\lim_{N \to \infty} \frac{f(N)}{g(N)} = 1$, and $f(N) \gtrsim g(N)$ is equivalent to $\lim_{N \to \infty} \frac{f(N)}{g(N)} \geq 1$.

## II. SYSTEM MODEL

In this work, we consider a MIMO-BC in which a BS equipped with $M$ antennas communicates with $N$ users, each equipped with $K$ antennas, where we assume that $K \leq M$. The channel between each user and the BS is modeled as a zero-mean circularly symmetric Gaussian matrix (Rayleigh fading). The received vector by user *k* can be written as

$$\mathbf{y}_k = \mathbf{H}_k \mathbf{x} + \mathbf{n}_k, \tag{1}$$

where $\mathbf{x} \in \mathbb{C}^{M \times 1}$ is the transmitted signal, $\mathbf{H}_k \in \mathbb{C}^{K \times M}$ is the channel matrix from the transmitter to the $k$th user, which is assumed to be perfectly known at the receiver side and partially known (or completely unknown) at the transmitter side, and $\mathbf{n}_k \in \mathbb{C}^{K \times 1} \sim \mathcal{CN}(\mathbf{0}, \mathbf{I}_K)$ is the noise vector at this receiver. We assume that the transmitter has an average power constraint $P$, i.e. $\mathbb{E}\left\{\text{Tr}(\mathbf{x}\mathbf{x}^{\text{H}})\right\} \leq P$. We consider a block fading model in which each $\mathbf{H}_k$ is constant for the duration of a frame. The frame itself is assumed to be long enough to allow communication at rates close to the capacity.

## III. ASYMPTOTIC ANALYSIS

### A. The average number of users send feedback to the BS

In this section, we define the amount of feedback as the average number of users who send feedback to the BS. It is assumed that the SNR ($P$) is fixed. In the following theorems, we provide the necessary and sufficient conditions in order to achieve $\lim_{N\to\infty} \frac{\mathcal{R}_S}{\mathcal{R}_{\text{Opt}}} = 1$ and $\lim_{N\to\infty} \mathcal{R}_{\text{Opt}} - \mathcal{R}_S = 0$, where $\mathcal{R}_{\text{Opt}}$ denotes the maximum achievable sum-rate in MIMO-BC, for any user selection strategy $S$, respectively:

**Theorem 1** *Consider a MIMO-BC with $N$ users ($N \to \infty$), which utilizes a fixed user selection strategy $S$. Let $\mathcal{N}_S$ be the number of users who send information to the BS in this strategy. Then, the necessary and sufficient condition to achieve $\lim_{N\to\infty} \frac{\mathcal{R}_S}{\mathcal{R}_{\text{Opt}}} = 1$ is having*

$$\mathbb{E}\{\mathcal{N}_S\} \sim \omega(1). \tag{2}$$

**Proof-** *Necessary Condition-* Let us denote $\mathcal{G}_S$ as the set of users who send information to the BS using strategy $S$. Define $p_S(k)$ as the probability that user $k$ belongs to $\mathcal{G}_S$. Since we consider a homogenous network, this probability is independent of $k$, and we denote it by $p_S$. Therefore, $\mathcal{N}_S = |\mathcal{G}_S|$ is a Binomial random variable with parameters $(N, p_S)$, and we have $\mathbb{E}\{\mathcal{N}_S\} = Np_S$.

Let us define

$$\mathcal{R}_1 = \mathbb{E}\left\{ \max_{\substack{\mathbf{Q}_n \\ \sum \text{Tr}(\mathbf{Q}_n)=P}} \ln\left|\mathbf{I}_M + \sum_{n=1}^{N} \mathbf{H}_n^H \mathbf{Q}_n \mathbf{H}_n\right| \,\bigg|\, \mathcal{A}_S \right\},$$

and

$$\mathcal{R}_2 = \mathbb{E}\left\{ \max_{\substack{\mathbf{Q}_n \\ \sum \text{Tr}(\mathbf{Q}_n)=P}} \ln\left|\mathbf{I}_M + \sum_{n=1}^{N} \mathbf{H}_n^H \mathbf{Q}_n \mathbf{H}_n\right| \,\bigg|\, \mathcal{A}_S^C \right\},$$

where $\mathcal{A}_S$ is the event that $|\mathcal{G}_S| = 0$, and $\mathcal{A}_S^C$ is the complement of $\mathcal{A}_S$. We have

$$\begin{aligned} \mathcal{R}_S &\leq \Pr\{\mathcal{A}_S\}\mathcal{R}_{\mathcal{A}_S}^{\text{NCSI}} + \Pr\{\mathcal{A}_S^C\}\mathcal{R}_2 \\ &= (1-p_S)^N \mathcal{R}_{\mathcal{A}_S}^{\text{NCSI}} + \left[1-(1-p_S)^N\right]\mathcal{R}_2, \end{aligned} \tag{3}$$

where $\mathcal{R}_S$ denotes the achievable sum-rate by the strategy $S$ and $\mathcal{R}^{\text{NCSI}}_{\mathcal{A}_S}$ stands for the sum-rate of MIMO-BC when no CSI is available at the BS, conditioned on $\mathcal{A}_S$. The above equation comes from the fact that with probability $\Pr\{\mathcal{A}_S\} = (1-p_S)^N$ no users send feedback to the BS and hence, the resulting sum-rate is upper-bounded by $\mathcal{R}^{\text{NCSI}}_{\mathcal{A}_S}$. Using (3) and having

$$\mathcal{R}_{\text{Opt}} = \Pr\{\mathcal{A}_S\}\mathcal{R}_1 + \Pr\{\mathcal{A}_S^C\}\mathcal{R}_2, \quad (4)$$

we can write

$$\mathcal{R}_{\text{Opt}} - \mathcal{R}_S \geq (1-p_S)^N (\mathcal{R}_1 - \mathcal{R}^{\text{NCSI}}_{\mathcal{A}_S}). \quad (5)$$

It can also be shown that

$$\mathcal{R}_1 \geq \mathbb{E}\left\{\ln\left(1 + P \max_{j,k} \|\mathbf{H}_{j,k}\|^2\right) \bigg| \mathcal{A}_S\right\}, \quad (6)$$

where $\mathbf{H}_{j,k}$ denotes the $j$th row of $\mathbf{H}_k$. The right hand side of (6) can be lower-bounded as,

$$\text{RH}(6) \geq \mathbb{E}\left\{\ln\left(1 + P \max_{j,k} \|\mathbf{H}_{j,k}\|^2\right) \bigg| \mathcal{A}_S, \mathscr{C}_t\right\} \Pr\{\mathscr{C}_t | \mathcal{A}_S\}, \quad (7)$$

where $\mathscr{C}_t$ is the event that $\max_{j,k} \|\mathbf{H}_{j,k}\|^2 > t$, for some chosen $t$. Hence,

$$\begin{aligned}
\text{RH}(6) &\geq \ln(1+Pt)\frac{\Pr\{\mathcal{A}_S, \mathscr{C}_t\}}{\Pr\{\mathcal{A}_S\}} \\
&\geq \ln(1+Pt)\frac{1 - \Pr\{\mathcal{A}_S^C\} - \Pr\{\mathscr{C}_t^C\}}{\Pr\{\mathcal{A}_S\}} \\
&= \ln(1+Pt)\left(1 - \frac{\Pr\{\mathscr{C}_t^C\}}{\Pr\{\mathcal{A}_S\}}\right),
\end{aligned} \quad (8)$$

where $\mathscr{C}_t^C$ is the complement of $\mathscr{C}_t$. $\Pr\{\mathscr{C}_t^C\}$ can be computed as

$$\begin{aligned}
\Pr\{\mathscr{C}_t^C\} &= \Pr\left\{\max_{j,k} \|\mathbf{H}_{j,k}\|^2 \leq t\right\} \\
&\stackrel{(a)}{=} \left(1 - \sum_{m=0}^{M-1} \frac{t^m}{m!} e^{-t}\right)^{NK},
\end{aligned} \quad (9)$$

where $(a)$ comes from the fact that $\|\mathbf{H}_{j,k}\|^2$ has chi-square distribution with $2M$ degrees of freedom [19]. Now, assume that

$$\mathbb{E}\{\mathcal{N}_S\} = Np_S \backsim \omega(1), \quad (10)$$

i.e., $Np_S \sim O(1)$. Choosing $t = \frac{\ln N}{2}$, from (9), we obtain

$$\Pr\{\mathscr{C}_t^C\} \sim e^{-\frac{K\sqrt{N}(\ln N)^{M-1}}{2^{M-1}(M-1)!}[1+o(1)]}. \quad (11)$$

Indeed, noting $\Pr\{\mathcal{A}_S\} = (1 - p_S)^N$ and $Np_S \sim O(1)$, we have

$$\Pr\{\mathcal{A}_S\} \sim \Theta(1). \tag{12}$$

Substituting (11) and (12) in (8) yields

$$\begin{aligned} \text{RH(6)} &\gtrsim \ln\left(1 + \frac{P}{2}\ln N\right)\left(1 - \Theta\left(e^{-\frac{K\sqrt{N}(\ln N)^{M-1}}{2^{M-1}(M-1)!}[1+o(1)]}\right)\right) \\ &\sim \ln\ln N + O(1). \end{aligned} \tag{13}$$

Indeed, using the fact that in a homogenous MIMO-BC (when the users' channels have the same statistical behavior) with no CSI at the transmitter, the maximum sum-rate is achieved by time-sharing between the users [9], we can write

$$\begin{aligned} \mathcal{R}_{\mathcal{A}_S}^{\text{NCSI}} &= \mathbb{E}_{\mathbf{H}_k|\mathcal{A}_S}\left\{\ln\left|\mathbf{I} + \frac{P}{M}\mathbf{H}_k\mathbf{H}_k^H\right|\,\Big|\,\mathcal{A}_S\right\} \\ &\leq K\mathbb{E}_{\mathbf{H}_k|\mathcal{A}_S}\left\{\ln\left(1 + \frac{P}{M}\|\mathbf{H}_k\|^2\right)\Big|\,\mathcal{A}_S\right\} \\ &\overset{(a)}{\leq} K\ln\left(1 + \frac{P}{M}\mathbb{E}_{\mathbf{H}_k|\mathcal{A}_S}\left\{\|\mathbf{H}_k\|^2\big|\,\mathcal{A}_S\right\}\right) \\ &\overset{(b)}{\leq} K\ln\left(1 + \frac{P}{M}\frac{\mathbb{E}_{\mathbf{H}_k}\{\|\mathbf{H}_k\|^2\}}{\Pr\{\mathcal{A}_S\}}\right) \\ &= K\ln\left(1 + \frac{PK}{\Pr\{\mathcal{A}_S\}}\right) \\ &\overset{(12)}{\sim} \Theta(1), \end{aligned} \tag{14}$$

where $(a)$ comes from the concavity of $\ln$ function and $(b)$ comes from the fact that $\mathbb{E}_{\mathbf{H}_k}\{\|\mathbf{H}_k\|^2\} \geq \mathbb{E}_{\mathbf{H}_k|\mathcal{A}_S}\{\|\mathbf{H}_k\|^2|\mathcal{A}_S\}\Pr\{\mathcal{A}_S\}$. Combining (6), (13), and (14), and substituting in (5), under the assumption of (10), we get

$$\begin{aligned} \mathcal{R}_{\text{Opt}} - \mathcal{R}_S &\geq \left(1 - \frac{O(1)}{N}\right)^N [\ln\ln N + O(1)] \\ &\sim e^{-O(1)}\ln\ln N. \\ \Rightarrow \frac{\mathcal{R}_S}{\mathcal{R}_{\text{Opt}}} &\leq 1 - \frac{e^{-O(1)}\ln\ln N}{\mathcal{R}_{\text{Opt}}}. \end{aligned} \tag{15}$$

As a result, noting that $\mathcal{R}_{\text{Opt}} \sim M\ln\ln N$ [14], we obtain

$$\mathbb{E}\{\mathcal{N}_S\} \not\sim \omega(1) \Rightarrow \lim_{N\to\infty} \frac{\mathcal{R}_S}{\mathcal{R}_{\text{Opt}}} \neq 1. \tag{16}$$

*Sufficient Condition*- Let us define the strategy $S$ as selecting $M$ users randomly among the following set:

$$\mathcal{G}_S = \{k | \lambda_{\max}(\mathbf{H}_k) > t\}, \tag{17}$$

where $\lambda_{\max}(\mathbf{H}_k)$ is the maximum singular value of $\mathbf{H}_k \mathbf{H}_k^H$, and $t$ is a threshold value. After selecting the users, the BS performs ZFBF, where the coordinates are chosen as the eigenvectors, corresponding to the maximum singular values of the selected users. In [20], it has been shown that for a $K \times M$ matrix $\mathbf{A}$, whose elements are i.i.d Gaussian, we have

$$p_S \triangleq \Pr\{\lambda_{\max}(\mathbf{A}) > t\} = \frac{t^{M+K-2} e^{-t}(1 + O(e^{-t} t^{-1}))}{\Gamma(M)\Gamma(K)}. \tag{18}$$

Hence,

$$\begin{aligned} \mathbb{E}\{\mathcal{N}_S\} &= N p_S \\ &= N \frac{t^{M+K-2} e^{-t}(1 + O(e^{-t} t^{-1}))}{\Gamma(M)\Gamma(K)}. \end{aligned} \tag{19}$$

Having $\mathbb{E}\{\mathcal{N}_S\} \sim \omega(1)$, yields,

$$t \sim \ln N + (M + K - 2) \ln \ln N - \omega(1). \tag{20}$$

Utilizing ZFBF at the BS, and defining

$$\mathcal{R}^* \triangleq M \mathbb{E}_{\boldsymbol{\mathcal{H}}} \left\{ \ln \left( 1 + \frac{P}{\text{Tr}\left\{ [\boldsymbol{\mathcal{H}}^{\text{H}} \boldsymbol{\mathcal{H}}]^{-1} \right\}} \right) \bigg| |\mathcal{G}_S| \geq M \right\},$$

we can write

$$\mathcal{R}_S \geq \mathcal{R}^* \Pr\{|\mathcal{G}_S| \geq M\}, \tag{21}$$

where $\boldsymbol{\mathcal{H}} = \left[ \mathbf{g}_{s_1,\max}^T | \mathbf{g}_{s_2,\max}^T | \cdots | \mathbf{g}_{s_m,\max}^T \right]^T$ in which $\mathbf{g}_{s_i,\max} = \sqrt{\lambda_{\max}(\mathbf{H}_{s_i})} \mathbf{V}_{s_i,\max}^H$, $i = 1, \cdots, m$ $(m \leq M)$, and $\mathbf{V}_{s_i,\max}$ is the eigenvector corresponding to maximum singular value of the $i$th selected user ($s_i$), and $m = \min(M, |\mathcal{G}_S|)$.

$\eta_S \triangleq \Pr\{|\mathcal{G}_S| \geq M\}$ can be computed as follows:

$$\begin{aligned} \eta_S &= 1 - \Pr\{|\mathcal{G}_S| < M\} \\ &= 1 - \sum_{m=0}^{M-1} \binom{N}{m} p_S^m (1 - p_S)^{N-m} \\ &\stackrel{(a)}{\geq} 1 - \sum_{m=0}^{M-1} \frac{(N p_S)^m}{m!} e^{-(N-m) p_S}, \end{aligned} \tag{22}$$

where $(a)$ results from the facts that $\binom{N}{m} \leq \frac{N^m}{m!}$ and $(1-p_S)^{N-m} \leq e^{-(N-m)p_S}$. Since $Np_S \sim \omega(1)$, we have $\eta_S \sim 1 - o(1)$.

Indeed, we can lower-bound $\mathcal{R}^*$ as

$$\mathcal{R}^* \geq M \ln P - M \mathbb{E}_{\mathcal{H}}\left\{ X(\mathcal{H}) | |\mathcal{G}_S| \geq M \right\}, \tag{23}$$

where $X(\mathcal{H}) \triangleq \ln\left(\operatorname{Tr}\left\{ \left[\mathcal{H}^{\mathrm{H}}\mathcal{H}\right]^{-1} \right\}\right)$. In [21], Appendix E, it has been shown that

$$\mathbb{E}_{\mathcal{H}}\left\{ X(\mathcal{H}) | |\mathcal{G}_S| \geq M \right\} \leq \ln \frac{M}{t} + (M-1)\ln(2M^2). \tag{24}$$

Using the above equation and (23) and selecting $t > \ln N$, yields,

$$\mathcal{R}^* \geq M \ln\left(\frac{P \ln N}{M}\right) - M(M-1)\ln(2M^2). \tag{25}$$

Substituting $\mathcal{R}^*$ and $\eta_S$ in (21), and having the fact that $\mathcal{R}_{\mathrm{Opt}} \sim M \ln \ln N$ [14], yields

$$\lim_{N \to \infty} \frac{\mathcal{R}_S}{\mathcal{R}_{\mathrm{Opt}}} = 1. \tag{26}$$

∎

**Theorem 2** *For any user selection strategy $S$, the necessary condition to achieve $\lim_{N \to \infty} \mathcal{R}_{\mathrm{Opt}} - \mathcal{R}_S = 0$ is having*

$$\mathbb{E}\{\mathcal{N}_S\} \sim \ln \ln \ln N + \omega(1). \tag{27}$$

**Proof -** Assume that

$$\mathbb{E}\{\mathcal{N}_S\} \nsim \ln \ln \ln N + \omega(1). \tag{28}$$

In other words, $\mathbb{E}\{\mathcal{N}_S\} \sim \ln \ln \ln N + O(1)$, or $\mathbb{E}\{\mathcal{N}_S\} < \ln \ln \ln N$. Similar to (5), we can write

$$\mathcal{R}_{\mathrm{Opt}} - \mathcal{R}_S \geq (1-p_S)^N [\mathcal{R}_1 - \mathcal{R}_{\mathcal{A}_S}^{\mathrm{NCSI}}]. \tag{29}$$

Following the same approach as in Theorem 1, under the assumption of (28), we can show that $\mathcal{R}_1 \gtrsim \ln \ln N + O(1)$, and $\mathcal{R}_{\mathcal{A}_S}^{\mathrm{NCSI}} \sim O(\ln \ln \ln N)$. Hence,

$$\begin{aligned}
\mathcal{R}_{\mathrm{Opt}} - \mathcal{R}_S &\geq (1-p_S)^N \left[\ln \ln N + O(\ln \ln \ln N)\right] \\
&\stackrel{(a)}{\sim} e^{-\mathbb{E}\{\mathcal{N}_S\}[1+O(p_S)]} \left[\ln \ln N + O(\ln \ln \ln N)\right] \\
&\stackrel{(b)}{\sim} e^{-(\mathbb{E}\{\mathcal{N}_S\} - \ln \ln \ln N)} \left[1 + o(1)\right].
\end{aligned} \tag{30}$$

(a) comes from the facts that $\mathbb{E}\{\mathcal{N}_S\} = Np_S$ and $\ln(1 - p_S) \sim p_S + O(p_S^2)$, and (b) results from writing $\ln \ln N$ as $e^{\ln \ln \ln N}$, noting that $e^{\mathbb{E}\{\mathcal{N}_S\}O(p_S)} \sim 1 + o(1)$. In the case of $\mathbb{E}\{\mathcal{N}_S\} \sim \ln \ln \ln N + O(1)$, we have $\text{RH}(30) \sim e^{-O(1)}\left[1 + o(1)\right]$. In the case of $\mathbb{E}\{\mathcal{N}_S\} < \ln \ln \ln N$, we have $\text{RH}(30) \sim \Upsilon\left[1 + o(1)\right]$, where $\Upsilon > 1$. As a result,

$$\mathbb{E}\{\mathcal{N}_S\} \nsim \ln \ln \ln N + \omega(1) \Rightarrow \lim_{N \to \infty} \mathcal{R}_{\text{Opt}} - \mathcal{R}_S \neq 0. \tag{31}$$

∎

**Theorem 3** *The sufficient condition to achieve $\lim_{N \to \infty} \mathcal{R}_{\text{opt}} - \mathcal{R}_S = 0$ is having*

$$\mathbb{E}\{\mathcal{N}_S\} \sim M \ln \ln \ln N + \omega(1). \tag{32}$$

**Proof -** Consider the Random Beam-Forming strategy, introduced in [14]. In this strategy, the BS randomly constructs $M$ orthogonal beams and transmits data to the users with the maximum SINR for each beam. Assuming each user's antenna as a separate user, we define the following set:

$$\mathcal{G}_{\text{RBF}}^{(m)} = \{k | \exists i, \quad \text{SINR}_{k,i}^{(m)} > t\}, \quad m = 1, \cdots, M, \tag{33}$$

where $\text{SINR}_{k,i}^{(m)}$ is the received SINR over the $i$th antenna of the $k$th user, for the $m$th transmitted beam. $\mathcal{G}_{\text{RBF}} = \bigcup_{m=1}^{M} \mathcal{G}_{\text{RBF}}^{(m)}$ is the set of users who send feedback to the BS. The achievable sum-rate by this scheme, denoted by $\mathcal{R}_{\text{RBF}}$, is lower-bounded as

$$\begin{aligned}
\mathcal{R}_{\text{RBF}} &\geq M \ln(1 + t) \Pr \left\{ \bigcap_{m=1}^{M} \mathscr{D}_m \right\} \\
&\geq M \ln(1 + t) \left( 1 - \sum_{m=1}^{M} \Pr\{\mathscr{D}_m^C\} \right),
\end{aligned} \tag{34}$$

where $\mathscr{D}_m$ is the event that $|\mathcal{G}_{\text{RBF}}^{(m)}| \geq 1$, and $\mathscr{D}_m^C$ is the complement of $\mathscr{D}_m$.

For a randomly chosen user $k$, we define

$$\begin{aligned}
p_k^{(m)} &\triangleq \Pr\{k \in \mathcal{G}_{\text{RBF}}^{(m)}\} \\
&= \Pr \left\{ \bigcup_{i=1}^{K} \mathscr{B}_{k,i}^{(m)} \right\} \\
&\leq \sum_{i=1}^{K} \eta_{k,i}^{(m)},
\end{aligned} \tag{35}$$

where $\mathscr{B}_{k,i}^{(m)}$ is the event that $\text{SINR}_{k,i}^{(m)} > t$ and $\eta_{k,i}^{(m)} \triangleq \Pr\{\mathscr{B}_{k,i}^{(m)}\}$, which is independent of $k$, $i$, $m$, and we denote it by $\eta$. Indeed, $p_k^{(m)}$ is independent of $k$, $m$, and is denoted by $p$. Hence, $p \leq K\eta$.

To evaluate the right hand side of (34), first we compute $\Pr\{\mathscr{D}_m^C\}$ as follows:

$$\begin{aligned}
\Pr\{\mathscr{D}_m^C\} &= (1-\eta)^{KN} \\
&\leq \left(1 - \frac{p}{K}\right)^{KN}.
\end{aligned} \quad (36)$$

Therefore,

$$\begin{aligned}
\text{RH}(34) &\geq M\ln(1+t)\left[1 - M\left(1 - \frac{p}{K}\right)^{KN}\right] \\
&\geq M\ln(1+t)[1 - Me^{-Np}].
\end{aligned} \quad (37)$$

Under the condition of (32), which implies that $\mathbb{E}\{\mathcal{N}_{\text{RBF}}\} \sim MNp \sim M\ln\ln\ln N + \omega(1)$, and knowing the fact that $\eta = \frac{e^{-Mt/P}}{(1+t)^{M-1}}$ [14] and writing $p$ as $p = T\eta$, where $T$ is a constant such that $1 \leq T \leq K$, we can write

$$\begin{aligned}
NT\frac{e^{-Mt/P}}{(1+t)^{M-1}} &\sim \ln\ln\ln N + \omega(1). \\
\Rightarrow t &\sim \frac{P}{M}\left[\ln N - (M-1)\ln\left(\frac{P}{M}\ln N\right) - \right. \\
&\left. \ln\ln\ln\ln N + \ln T - \omega\left(\frac{1}{\ln\ln\ln N}\right)\right].
\end{aligned} \quad (38)$$

Substituting $t$ in (37) yields

$$\begin{aligned}
\mathcal{R}_{\text{RBF}} &\geq M\ln\left(1 + \frac{P}{M}\ln N + O(\ln\ln N)\right) \times \\
&\quad \left(1 - Me^{-Np}\right).
\end{aligned} \quad (39)$$

Using the above equation and having the facts that $\mathcal{R}_{\text{Opt}} \sim M\ln\left(1 + \frac{P}{M}\ln N + O(\ln\ln N)\right)$ [14], and $\mathbb{E}\{\mathcal{N}_{\text{RBF}}\} \sim M\ln\ln\ln N + \omega(1)$, we have

$$\begin{aligned}
\mathcal{R}_{\text{Opt}} - \mathcal{R}_{\text{RBF}} &\leq O\left(\frac{\ln\ln N}{\ln N}\right) + M^2 e^{-(\frac{\mathbb{E}\{\mathcal{N}_{\text{RBF}}\}}{M} - \ln\ln\ln N)}[1 + o(1)] \\
&\sim o(1).
\end{aligned} \quad (40)$$

Consequently, $\lim_{N\to\infty} \mathcal{R}_{\text{Opt}} - \mathcal{R}_{\text{RBF}} = 0$.

∎

## B. Amount of bits fed back to the BS

In this section, we study the minimum amount of feedback required at the BS, in terms of number of bits [2], in order to achieve the maximum sum-rate capacity. It is assumed that the SNR ($P$) is fixed and the number of bits fed back by each user is an integer.

**Theorem 4** *The necessary and sufficient condition to achieve* $\lim_{N \to \infty} \frac{\mathcal{R}_S}{\mathcal{R}_{\text{Opt}}} = 1$ *for any user selection strategy $S$ is having*

$$\mathbb{E}\{\mathcal{F}_S\} \sim \omega(1), \tag{41}$$

*where $\mathcal{F}_S$ is the total number of bits fed back to the BS.*

**Proof-** *Necessary condition-* The proof of the necessary condition easily follows from Theorem 1, and the fact that the number of bits fed back by each user is an integer.

*Sufficient Condition-* Consider the Random Beam-Forming scheme. Given any function $f(N) \triangleq \mathbb{E}\{\mathcal{N}_S\} \sim \omega(1)$, we set the threshold $t$ as the solution to the following equation:

$$\frac{e^{-Mt/P}}{(1+t)^{M-1}} = \frac{f(N)}{MNT}, \tag{42}$$

where $T$ is a constant between 1 and $K$. By selecting $t$ as the above equation, using the same approach as in the proof of Theorem 3, it can be shown that $\lim_{N \to \infty} \frac{\mathcal{R}_S}{\mathcal{R}_{\text{Opt}}} = 1$. Since the users in $\mathcal{G}_{\text{RBF}}^{(m)}$ only need to send the index $m$ to the BS, the total amount of feedback bits is equal to $\lceil \log_2(M) \rceil f(N) \sim \omega(1)$. Consequently, it is possible to achieve $\lim_{N \to \infty} \frac{\mathcal{R}_S}{\mathcal{R}_{\text{Opt}}} = 1$, with the average number of feedback bits scaling as $\omega(1)$.

∎

**Theorem 5** *The necessary and sufficient condition to achieve* $\lim_{N \to \infty} \mathcal{R}_{\text{Opt}} - \mathcal{R}_S = 0$ *is having*

$$\mathbb{E}\{\mathcal{F}_S\} \sim \Theta(\ln \ln \ln N) + \omega(1). \tag{43}$$

**Proof-** The proof follows from Theorems 2 and 3, with the same approach as that of Theorem 4.

∎

---

[2] In fact, it is more precise to express the amount of feedback in terms of *binits*, as it is assumed that the users who do not send any information to the BS do not contribute to the total amount of feedback.

*Remark 1-* From the above theorems, it follows that the Random Beam-forming scheme is optimum in the fixed SNR regime, in the sense of achieving the maximum sum-rate with the minimum required amount of feedback.

*Remark 2-* Using the conventional ZFBF (with the user selection algorithm as in the proof of the sufficient condition in Theorem 1), assuming that the selected users quantize the eigenvectors corresponding to their maximum singular values and feed back the quantization indices to the BS, from [22], it can be shown that

$$\mathcal{R}_{\text{ZF}} - \mathcal{R}_{\text{ZF}}^{\text{Q}} \leq M \ln \left(1 + P\gamma(\ln N)2^{-\frac{B}{M-1}}\right), \tag{44}$$

where $\mathcal{R}_{\text{ZF}}$ denotes the achievable sum-rate of ZFBF when the BS has perfect CSI from all the selected users, $\mathcal{R}_{\text{ZF}}^{\text{Q}}$ is the achievable sum-rate when the BS only has the quantization indices of the selected users' channels, $B$ is the number of quantized bits for each selected user, and $\gamma$ is a constant depending on the quantization method, which is shown to be lower-bounded by $\frac{M-1}{M}$ [22]. From the above equation, it follows that in order to achieve $\lim_{N\to\infty} \frac{\mathcal{R}_{\text{ZF}}^{Q}}{\mathcal{R}_{\text{Opt}}} = 1$, we must have $B \gtrsim (M-1)\ln \ln N + o(\ln \ln N)$, and in order to achieve $\lim_{N\to\infty} \mathcal{R}_{\text{Opt}} - \mathcal{R}_{\text{ZF}}^{Q} = 0$, the condition $B \sim (M-1)\ln \ln N + \omega(1)$ must be satisfied. In other words, the minimum required amount of feedback to achieve the maximum sum-rate must scale at least as $\ln \ln N$. This implies that although the proposed user selection algorithm in Theorem 1, along with utilizing ZFBF, is shown to be optimal in terms of the average number of users who send feedback to the BS, in terms of the average number of feedback bits, it is not optimal.

## C. Variable SNR Scenario

In the previous section, the SNR ($P$) is assumed to be fixed. In this section, we study the scaling law of the minimum amount of feedback in order to achieve the maximum sum-rate, when the SNR itself is a function of $N$. To this end, we consider two special regimes of *low SNR* and *high SNR*. Since achieving the optimum sum-rate requires the square magnitudes of the selected coordinates to behave as $\ln N$, the effective SNR of the selected links scales as $P \ln N$. Hence, *low SNR* and *high SNR* regimes are defined by the regions of $P \ln N \sim o(1)$ and $P \ln N \sim \omega(1)$, respectively.

*1) Low SNR Regime:* In this regime, it can be shown that [23]

$$\mathcal{R}_{\text{opt}} \sim P\mathbb{E}\{\eta_{\max}\}, \tag{45}$$

where $\eta_{\max} \triangleq \max_k \lambda_{\max}(\mathbf{H}_k)$. In other words, the optimum strategy requires the BS to perform beam-forming on the eigenvector corresponding to the maximum largest eigenvalue. Having the fact that $\mathbb{E}\{\eta_{\max}\} \sim \ln N$ [20], it follows that in the low SNR regime, as $\mathcal{R}_{\text{opt}} \sim P \ln N \sim o(1)$, the achievability of the optimum sum-rate for a given strategy $S$ is defined by $\lim_{N \to \infty} \frac{\mathcal{R}_S}{\mathcal{R}_{\text{opt}}} = 1$.

**Theorem 6** *The necessary and sufficient condition in order to achieve the optimum sum-rate throughput in the low SNR regime is:*

$$\mathbb{E}\{\mathcal{N}_S\} \sim \omega(1),$$

*and*

$$\mathbb{E}\{\mathcal{F}_S\} \sim \omega(1).$$

**Proof** - Following the approach of Theorem 1 and using the equations (5), (8), (9), and (14), we have

$$\mathcal{R}_{\text{Opt}} - \mathcal{R}_S \geq (1 - p_S)^N \left( \mathcal{R}_1 - \mathcal{R}_{\mathcal{A}_S}^{\text{NCSI}} \right), \tag{46}$$

$$\mathcal{R}_1 \geq \ln(1 + Pt) \left( 1 - \frac{\left(1 - \sum_{m=0}^{M-1} \frac{t^m}{m!} e^{-t}\right)^{NK}}{(1 - p_S)^N} \right)$$

$$\stackrel{(a)}{\sim} Pt \left( 1 - \frac{\left(1 - \sum_{m=0}^{M-1} \frac{t^m}{m!} e^{-t}\right)^{NK}}{(1 - p_S)^N} \right), \tag{47}$$

and

$$\mathcal{R}_{\mathcal{A}_S}^{\text{NCSI}} \leq K \ln \left( 1 + \frac{PK}{(1 - p_S)^N} \right). \tag{48}$$

$(a)$ comes from the low-SNR assumption and the fact that for $x \ll 1$, $\ln(1+x) \approx x$. Under the assumption of $\mathbb{E}\{\mathcal{N}_S\} = N p_S \sim O(1)$ and choosing $t = \frac{\ln N}{2}$, we have $\mathcal{R}_1 \sim \frac{P \ln N}{2}$ and $\mathcal{R}_{\mathcal{A}_S}^{\text{NCSI}} \sim \Theta(P)$. Noting that $\mathcal{R}_{\text{Opt}} \sim P \ln N$, we can write

$$\frac{\mathcal{R}_S}{\mathcal{R}_{\text{Opt}}} \leq 1 - (1 - p_S)^N \frac{\mathcal{R}_1 - \mathcal{R}_{\mathcal{A}_S}^{\text{NCSI}}}{\mathcal{R}_{\text{Opt}}}$$

$$\sim 1 - \Theta(1). \tag{49}$$

As a result,

$$\mathbb{E}\{\mathcal{N}_S\} \not\sim \omega(1) \Rightarrow \lim_{N \to \infty} \frac{\mathcal{R}_S}{\mathcal{R}_{\text{Opt}}} < 1. \tag{50}$$

The necessity of $\mathbb{E}\{\mathcal{F}_S\} \sim \omega(1)$ directly follows from the above equation.

*Sufficient condition* - In this part, we prove that for any given function $f(N) \sim \omega(1)$, one can achieve the maximum sum-rate such that $\mathbb{E}\{\mathcal{N}_S\} \leq f(N)$ and $\mathbb{E}\{\mathcal{F}_S\} \leq f(N)$. Assume that the users in the following set:

$$\mathcal{G}_S \triangleq \{k | \lambda_{\max}(\mathbf{H}_k) > t\}, \tag{51}$$

where

$$t \triangleq \max\left(\ln N + (M + K - 2)\ln\ln N - \frac{1}{2}\ln f(N), \ln N\right), \tag{52}$$

quantize the eigenvector corresponding to their maximum singular value, using a quantization code book $\mathcal{W}$, which consists of $L = 2^{\frac{\sqrt{f(N)}}{2}}$ randomly selected unit vectors in the $M$-dimensional space (Random Vector Quantization (RVQ)). The BS selects one of the users in $\mathcal{G}_S$ at random and serves this user, performing beam-forming on the direction of its quantized eigenvector. The achievable sum-rate of this scheme can be lower-bounded as

$$\begin{aligned} \mathcal{R}_S &\geq \mathbb{E}\left\{\ln\left(1 + Pt|\mathbf{\Phi}^H\widehat{\mathbf{\Phi}}|^2\right)\right\}\left[1 - (1-p_S)^N\right] \\ &\approx Pt\,\mathbb{E}\left\{|\mathbf{\Phi}^H\widehat{\mathbf{\Phi}}|^2\right\}\left[1 - (1-p_S)^N\right] \\ &\stackrel{(a)}{\geq} Pt\,\mathbb{E}\left\{|\mathbf{\Phi}^H\widehat{\mathbf{\Phi}}|^2\right\}\left[1 - e^{-Np_S}\right], \end{aligned} \tag{53}$$

where $p_S \triangleq \Pr\{k \in \mathcal{G}_S\}$ for a randomly chosen $k$, $\mathbf{\Phi}$ denotes the eigenvector corresponding to the maximum singular value of the selected user, and $\widehat{\mathbf{\Phi}}$ denotes the quantized version of $\mathbf{\Phi}$. $(a)$ comes from the fact that $(1-p_S)^N \leq e^{-Np_S}$. Using (18), we can write

$$\begin{aligned} p_S &\sim \frac{t^{M+K-2}e^{-t}}{\Gamma(M)\Gamma(K)}\left[1 + O(e^{-t}t^{-1})\right] \\ &\stackrel{(a)}{\sim} \min\left(\frac{\sqrt{f(N)}}{N}, \frac{(\ln N)^{M+K-2}}{N}\right) \\ \Rightarrow e^{-Np_S} &\sim e^{-\min\left(\sqrt{f(N)}, (\ln N)^{M+K-2}\right)}, \end{aligned} \tag{54}$$

where $(a)$ comes from (52). We have

$$\begin{aligned} \theta &\triangleq |\mathbf{\Phi}^H\widehat{\mathbf{\Phi}}|^2 \\ &= \max_{\substack{\mathbf{c}_l \\ \mathbf{c}_l \in \mathcal{W}}} |\mathbf{\Phi}^H\mathbf{c}_l|^2. \end{aligned} \tag{55}$$

From [21], Appendix C, it follows that the pdf of $\theta_l \triangleq |\mathbf{\Phi}^H \mathbf{c}_l|^2$ is obtained from

$$f_{\theta_l}(\theta_l) = (M-1)(1-\theta_l)^{M-2}, \quad 0 \leq \theta_l \leq 1. \tag{56}$$

Hence,

$$\begin{aligned}
F_\theta(\theta) &= [F_{\theta_l}(\theta)]^L \\
&= \left[1 - (1-\theta)^{M-1}\right]^L.
\end{aligned} \tag{57}$$

From the above equation, $\mathbb{E}\{\theta\}$ can be lower-bounded as

$$\begin{aligned}
\mathbb{E}\{\theta\} &= \int_0^1 \theta f_\theta(\theta) \mathrm{d}\theta \\
&= \int_0^1 (1 - F_\theta(\theta)) \mathrm{d}\theta \\
&= \int_0^1 \left(1 - \left[1 - (1-\theta)^{M-1}\right]^L\right) \mathrm{d}\theta \\
&= \int_0^1 \left(1 - \left[1 - \mu^{M-1}\right]^L\right) \mathrm{d}\mu \\
&\stackrel{(a)}{\geq} 1 - \int_0^1 e^{-L\mu^{M-1}} \mathrm{d}\mu \\
&\stackrel{(b)}{=} 1 - \frac{L^{-\frac{1}{M-1}}}{M-1} \int_0^L u^{\frac{2-M}{M-1}} e^{-u} \mathrm{d}u \\
&\stackrel{(c)}{\geq} 1 - \frac{L^{-\frac{1}{M-1}}}{M-1} \left[\int_0^1 u^{\frac{2-M}{M-1}} \mathrm{d}u + \int_1^\infty e^{-u} \mathrm{d}u\right] \\
&= 1 - L^{-\frac{1}{M-1}} \left(1 + \frac{e^{-1}}{M-1}\right) \\
&\stackrel{(d)}{=} 1 - 2^{-\frac{\sqrt{f(N)}}{2(M-1)}} \left(1 + \frac{e^{-1}}{M-1}\right).
\end{aligned} \tag{58}$$

In the above equation, $(a)$ comes from the fact that $\left[1 - \mu^{M-1}\right]^L \leq e^{-L\mu^{M-1}}$, $(b)$ results from the change of variable $u = L\mu^{M-1}$. $(c)$ comes from the fact that as $M \geq 2$, $\frac{2-M}{M-1} \leq 0$, and as a result, for $u \geq 1$, $u^{\frac{2-M}{M-1}} \leq 1$. $(d)$ follows from the definition of $L$ as $2^{\frac{\sqrt{f(N)}}{2}}$. Combining (45), (52), (53), (54), and (58), and the fact that $\mathbb{E}\{\eta_{\max}\} \sim \ln N + O(\ln \ln N)$ [20], yields,

$$\begin{aligned}
\lim_{N \to \infty} \frac{\mathcal{R}_S}{\mathcal{R}_{\mathrm{Opt}}} &= \lim_{N \to \infty} \frac{Pt \left[1 - 2^{-\frac{\sqrt{f(N)}}{2(M-1)}} \left(1 + \frac{e^{-1}}{M-1}\right)\right] \left[1 - e^{-\min\left(\sqrt{f(N)}, (\ln N)^{M+K-2}\right)}\right]}{P \mathbb{E}\{\eta_{\max}\}} \\
&= 1.
\end{aligned} \tag{59}$$

Moreover, we have

$$\begin{aligned}
\mathbb{E}\{\mathcal{N}_S\} &= Np_S \\
&\lesssim \sqrt{f(N)} \\
&< f(N),
\end{aligned} \quad (60)$$

and

$$\begin{aligned}
\mathbb{E}\{\mathcal{F}_S\} &= \mathbb{E}\{\mathcal{N}_S\}\log_2(L) \\
&\lesssim \sqrt{f(N)}\,\frac{\sqrt{f(N)}}{2} \\
&< f(N),
\end{aligned} \quad (61)$$

which completes the proof of Theorem 6.

∎

*2) High SNR Regime:* The sum-rate capacity in this regime can be written as [14],

$$\mathcal{R}_{\text{Opt}} \sim M \ln\left(\frac{P}{M}\ln N + O(P\ln\ln N)\right). \quad (62)$$

**Theorem 7** *i) The necessary condition to achieve $\lim_{N\to\infty}\frac{\mathcal{R}_S}{\mathcal{R}_{\text{Opt}}} = 1$ in the case of $K < M$, and also $K = M$ and SNR regime of $\ln P \sim O(\ln\ln N)$, is having $\mathbb{E}\{\mathcal{N}_S\} \sim \omega(1)$. ii) in the case of $K = M$, and the regime of $\ln P \sim \omega(\ln\ln N)$, it is possible to achieve $\lim_{N\to\infty}\frac{\mathcal{R}_S}{\mathcal{R}_{\text{Opt}}} = 1$ without any CSI at the BS.*

**Proof -** *Proof of i)*: Similar to the proof of Theorem 1, we can write

$$\mathcal{R}_{\text{Opt}} - \mathcal{R}_S \geq (1-p_S)^N(\mathcal{R}_1 - \mathcal{R}_{\mathcal{A}_S}^{\text{NCSI}}). \quad (63)$$

From [20], $\mathcal{R}_1$ can be lower bounded as

$$\mathcal{R}_1 \geq \mathbb{E}\left\{\sum_{j=1}^{M}\log(1+\frac{P}{M}\sigma_j^2)\bigg|\mathcal{A}_S\right\}, \quad (64)$$

where

$$\begin{aligned}
\sigma_j^2 = \max_k \max_{\mathbf{x}} \ &\mathbf{x}^H \mathbf{H}_k^H \mathbf{H}_k \mathbf{x} \\
\text{s.t.} \quad &\mathbf{x}^H \mathbf{x} = 1 \\
&\mathbf{\Xi}_j^H \mathbf{x} = 0,
\end{aligned} \quad (65)$$

and $\Xi_j \triangleq [\mathbf{v}_1|\cdots|\mathbf{v}_{j-1}]$, in which $\mathbf{v}_i$, $i = 1, \cdots, j-1$, is the optimizing parameter $\mathbf{x}$, in the maximization of $\sigma_i^2$. In other words, the maximizing parameter $\mathbf{x}$ is found in the null space of the previously selected coordinates. Defining $\mathscr{C}_t \triangleq \left\{ \bigcap_{j=1}^{M} \left( \sigma_j^2 > t \right) \right\}$, similar to (8), we can write

$$\begin{aligned} \mathcal{R}_1 &\geq M \ln(1 + \frac{P}{M}t) \left( 1 - \frac{\Pr\{\mathscr{C}_t^C\}}{\Pr\{\mathcal{A}_S\}} \right) \\ &\stackrel{(a)}{\geq} M \ln(1 + \frac{P}{M}t) \left( 1 - \frac{\sum_{j=1}^{M} \Pr\{\sigma_j^2 \leq t\}}{\Pr\{\mathcal{A}_S\}} \right), \end{aligned} \qquad (66)$$

where $(a)$ comes from the union bound for the probability. From [20], Lemma 3, we have

$$\Pr\{\sigma_j^2 \leq t\} \leq \sum_{i=N-j+1}^{N} \binom{N}{i} G_{K,M-j+1}(t)^i \left[ 1 - G_{K,M-j+1}(t) \right]^{N-i}, \qquad (67)$$

where $G_{n,m}(t)$ is defined in [20], Lemma 1.

Setting $t = \frac{\ln N}{2}$, and using the result of [20], Appendix IV, on the asymptotic behavior of $G_{n,m}(t)$ for large $t$, we have

$$\begin{aligned} \Pr\left\{ \sigma_j^2 \leq \frac{\ln N}{2} \right\} &\leq \sum_{i=N-j+1}^{N} \binom{N}{i} \left[ 1 - \Theta\left( \frac{(\ln N)^{M+K-j-1}}{\sqrt{N}} \right) \right]^i \left[ \Theta\left( \frac{(\ln N)^{M+K-j-1}}{\sqrt{N}} \right) \right]^{N-i} \\ &\leq N^{j-1} e^{-\Theta\left(\sqrt{N}(\ln N)^{M+K-j-1}\right)} \\ &\sim o\left( N^{j-1} e^{-\sqrt{N}} \right). \end{aligned} \qquad (68)$$

Substituting in (66), we obtain

$$\mathcal{R}_1 \geq M \ln \left( 1 + \frac{P \ln N}{2M} \right) \left( 1 - \frac{o(N^{M-1} e^{-\sqrt{N}})}{\Pr\{\mathcal{A}_S\}} \right). \qquad (69)$$

Assuming $Np_S \not\sim \omega(1)$, noting that $\Pr\{\mathcal{A}_S\} = (1-p_S)^N$, incurs $\Pr\{\mathcal{A}_S\} \sim \Theta(1)$, which yields

$$\mathcal{R}_1 \geq M \ln \left( 1 + \frac{P \ln N}{2M} \right) \left( 1 - o(N^{M-1} e^{-\sqrt{N}}) \right). \qquad (70)$$

Moreover, using (14), under the condition of $Np_S \not\sim \omega(1)$, we have

$$\mathcal{R}_{\mathcal{A}_S}^{\text{NCSI}} \lesssim K \ln \left( \frac{P}{M} \right) + \Theta(1). \qquad (71)$$

Substituting in (63), yields

$$\mathcal{R}_{\text{Opt}} - \mathcal{R}_{\text{S}} \gtrsim (1-p_S)^N \left[ (M-K) \ln \left( \frac{P}{M} \ln N \right) + K \ln \ln N \right]. \qquad (72)$$

In the case of $K < M$, from the above equation and noting $\mathcal{R}_{\text{Opt}} \sim M \ln\left(\frac{P}{M} \ln N\right)$, it follows that

$$\frac{\mathcal{R}_{\text{S}}}{\mathcal{R}_{\text{Opt}}} \lesssim 1 - \frac{(1-p_S)^N (M-K)}{M}. \tag{73}$$

Hence, having $Np_S \asymp \omega(1)$ results in

$$\lim_{N,P \to \infty} \frac{\mathcal{R}_{\text{S}}}{\mathcal{R}_{\text{Opt}}} \neq 1. \tag{74}$$

Indeed, in the case of $K = M$, similar to (73), we can write

$$\frac{\mathcal{R}_{\text{S}}}{\mathcal{R}_{\text{Opt}}} \lesssim 1 - \frac{(1-p_S)^N \ln \ln N}{\ln P + \ln \ln N}. \tag{75}$$

Therefore, for the regime of $\ln P \sim O(\ln \ln N)$, having $Np_S \asymp \omega(1)$ incurs $\lim_{N \to \infty} \frac{\mathcal{R}_{\text{S}}}{\mathcal{R}_{\text{Opt}}} \neq 1$.

*Proof of ii)*: In the case of $K = M$ and $\ln P \sim \omega(\ln \ln N)$, assume that no CSI is available at the BS. In this case, the best strategy, as mentioned earlier, is time-sharing between the users. The achievable sum-rate in this case can be written as

$$\begin{aligned}\mathcal{R}_{\text{S}} &= \mathbb{E}\left\{\ln\left|\mathbf{I} + \frac{P}{M}\mathbf{H}_k\mathbf{H}_k^H\right|\right\} \\ &\approx M \ln(\frac{P}{M}) + \mathbb{E}\left\{\ln\left|\mathbf{H}_k\mathbf{H}_k^H\right|\right\} \\ &\sim M \ln P + \Theta(1). \end{aligned} \tag{76}$$

As a result,

$$\begin{aligned}\lim_{N \to \infty} \frac{\mathcal{R}_{\text{S}}}{\mathcal{R}_{\text{Opt}}} &= \lim_{N \to \infty} \frac{M \ln P}{M \ln P + M \ln \ln N} \\ &= 1. \end{aligned} \tag{77}$$

■

**Theorem 8** *The necessary condition to achieve $\lim_{N \to \infty} \mathcal{R}_{\text{Opt}} - \mathcal{R}_S = 0$ in the case of $K = M$ is having*

$$\mathbb{E}\{\mathcal{N}_S\} \sim \ln \ln \ln N + \omega(1), \tag{78}$$

*and in the case of $K < M$ is having*

$$\mathbb{E}\{\mathcal{N}_S\} \sim \ln \ln(P \ln N) + \omega(1), \tag{79}$$

*for the values of $P$ satisfying $\ln \ln(P \ln N) \sim o(N)$.*

**Proof** - The proof easily follows from (72) and the approach used in the proof of Theorem 2.

∎

Theorem 8 implies that in the case of $K = M$, the average number of users sending feedback to the BS does not need to grow with the SNR [3]. In the case of $K < M$, writing $\ln\ln(P \ln N)$ as $\ln\ln\ln N + \ln\left(1 + \frac{\ln P}{\ln\ln N}\right)$, it turns out that for the values of $P$ such that $\ln P \sim O(\ln\ln N)$, the condition $\mathbb{E}\{\mathcal{N}_S\} \sim \ln\ln(P \ln N) + \omega(1)$ is equivalent to $\mathbb{E}\{\mathcal{N}_S\} \sim \ln\ln\ln N + \omega(1)$, which implies that $\mathbb{E}\{\mathcal{N}_S\}$ does not need to grow with SNR. Moreover, for the values of $P$ satisfying $\ln P \sim \omega(\ln\ln N)$, the condition (79) reduces to $\mathbb{E}\{\mathcal{N}_S\} \sim \ln\ln P + \omega(1)$, which incurs that the average number of users sending feedback to the BS must grow at least double logarithmic with SNR.

In the previous section, we have observed that the Random Beam-forming scheme introduced in [14] is asymptotically optimal in the sense of achieving the maximum sum-rate with the minimum order of the required amount of feedback, in the fixed SNR regime. The question here is for what ranges of SNR this optimality holds. The following theorem answers this question:

**Theorem 9** *The necessary and sufficient condition to achieve* $\lim_{N\to\infty} \mathcal{R}_{\text{Opt}} - \mathcal{R}_{\text{RBF}} = 0$ *is having* [4]

$$\ln P \not\sim \Omega(\ln N). \tag{80}$$

**Proof** - *Necessary condition* - The sum-rate throughput of Random Beam-forming scheme can be upper-bounded as

$$\begin{aligned}
\mathcal{R}_{\text{RBF}} &= \mathbb{E}\left\{\sum_{m=1}^{M} \ln\left(1 + \text{SINR}_{\max}^{(m)}\right)\right\} \\
&\leq M \ln\left(1 + \mathbb{E}\{\text{SINR}_{\max}^{(m)}\}\right),
\end{aligned} \tag{81}$$

---

[3]This statement will be made rigorous in the proof of Theorem 11.

[4]It is assumed that each received antenna is treated as a separate user.

where $\text{SINR}_{\max}^{(m)}$ denotes the maximum received SINR over the $m$th transmitted beam. Defining $X_{\max} \triangleq \text{SINR}_{\max}^{(m)}$, for all values of $t$, we can write

$$\begin{aligned}
\mathbb{E}\{X_{\max}\} &= \int_0^\infty x f_{X_{\max}}(x) \mathrm{d}x \\
&= \int_0^\infty [1 - F_{X_{\max}}(x)] \, \mathrm{d}x \\
&\leq t + \int_t^\infty [1 - F_{X_{\max}}(x)] \, \mathrm{d}x, \quad t \geq 0.
\end{aligned} \quad (82)$$

Having the fact that $F_X(x) = 1 - \frac{e^{-\frac{Mx}{P}}}{(1+x)^{M-1}}$ [14], where $X \triangleq \text{SINR}_{i,k}^{(m)}$, we can write

$$\mathbb{E}\{X_{\max}\} \leq t + \int_t^\infty \left[1 - \left(1 - \frac{e^{-\frac{Mx}{P}}}{(1+x)^{M-1}}\right)^{NK}\right] \mathrm{d}x. \quad (83)$$

Assuming that $\ln P \sim \Omega(\ln N)$, i.e., $\lim_{N \to \infty} \frac{\ln P}{\ln N} = c$, where $c > 0$, we define

$$t \triangleq \begin{cases} \frac{P}{M}[\ln N - \frac{1}{2} \ln P], & c < 1; \\ \frac{P}{2M} \ln N, & c \geq 1. \end{cases} \quad (84)$$

Substituting $t$ in (83) yields,

$$\begin{aligned}
\mathbb{E}\{X_{\max}\} &\lesssim t + \int_t^\infty \left(1 - \exp\left\{-\frac{NKe^{-\frac{Mx}{P}}}{(1+x)^{M-1}}\left[1 + O\left(\frac{e^{-\frac{Mx}{P}}}{(1+x)^{M-1}}\right)\right]\right\}\right) \mathrm{d}x \\
&\overset{(a)}{\leq} t + \int_t^\infty \frac{NKe^{-\frac{Mx}{P}}}{(1+x)^{M-1}} \left[1 + O\left(\frac{e^{-\frac{Mx}{P}}}{(1+x)^{M-1}}\right)\right] \mathrm{d}x \\
&\leq t + \int_t^\infty \frac{NKe^{-\frac{Mx}{P}}}{(1+x)^{M-1}} \mathrm{d}x \left[1 + O\left(\frac{e^{-\frac{Mt}{P}}}{(1+t)^{M-1}}\right)\right] \\
&\overset{(b)}{\sim} t + \int_t^\infty \frac{NKe^{-\frac{Mx}{P}}}{(1+x)^{M-1}} \mathrm{d}x \left[1 + O\left(\frac{1}{\sqrt{N}}\right)\right] \\
&\overset{(c)}{\leq} t + \int_t^\infty \frac{NKe^{-\frac{Mx}{P}}}{(\frac{P}{M})^{M-1}} \mathrm{d}x \left[1 + O\left(\frac{1}{\sqrt{N}}\right)\right] \\
&\leq t + \left(\frac{P}{M}\right)^{2-M} NKe^{-\frac{Mt}{P}} \left[1 + O\left(\frac{1}{\sqrt{N}}\right)\right] \\
&\overset{(d)}{\sim} t + NKe^{-\frac{Mt}{P}} \left[1 + O\left(\frac{1}{\sqrt{N}}\right)\right] \\
&\leq \begin{cases} \frac{P}{M}[\ln N - \frac{1}{2} \ln P]\left[1 + O\left(\frac{1}{\sqrt{P}}\right)\right], & c < 1; \\ \frac{P}{2M} \ln N \left[1 + O\left(\frac{\sqrt{N}}{P}\right)\right], & c \geq 1, \end{cases}
\end{aligned} \quad (85)$$

where $(a)$ comes from the fact that $1 - e^{-x} \leq x$, $\forall x$, $(b)$ comes from the fact that $t \geq \frac{P}{2M} \ln N$ (from (84)), which incurs $\frac{e^{-\frac{Mt}{P}}}{(1+t)^{M-1}} \leq \frac{1}{\sqrt{N}}$, $(c)$ comes from the fact that since $t \geq \frac{P}{2M} \ln N$, for $x > t$, we have $1 + x > \frac{P}{M}$, and $(d)$ comes from the fact that $M \geq 2$ and as a result $\left(\frac{P}{M}\right)^{2-M} \leq 1$. Noting that $\mathcal{R}_{\text{Opt}} \sim M \ln \left(\frac{P \ln N}{M}\right)$, and using (81), (83), (84), and (85), we can write

$$\mathcal{R}_{\text{Opt}} - \mathcal{R}_{\text{RBF}} \geq \begin{cases} -\ln\left(1 - \frac{\ln P}{2 \ln N}\right) + O\left(\frac{1}{\sqrt{P}}\right), & c < 1; \\ \ln(2) - \ln\left[1 + O\left(\frac{\sqrt{N}}{P}\right)\right], & c \geq 1. \end{cases} \quad (86)$$

Noting that in the case of $c \geq 1$, $\frac{\sqrt{N}}{P} \sim o(1)$, it follows from the above equation that

$$\ln P \sim \Omega(\ln N) \Rightarrow \lim_{N \to \infty} \mathcal{R}_{\text{Opt}} - \mathcal{R}_{\text{RBF}} \neq 0. \quad (87)$$

*Sufficient condition* - Assume that $\ln P \nsim \Omega(\ln N)$. $\mathcal{R}_{\text{RBF}}$ can be lower-bounded as

$$\begin{aligned} \mathcal{R}_{\text{RBF}} &\geq M \ln(1+t) \Pr\left\{\text{SINR}_{\max}^{(1)} > t, \cdots, \text{SINR}_{\max}^{(M)} > t\right\} \\ &\geq M \ln(1+t) \left[1 - \sum_{m=1}^{M} \Pr\left\{\text{SINR}_{\max}^{(m)} \leq t\right\}\right] \\ &= M \ln(1+t) \left[1 - M(1-\eta)^{NK}\right] \\ &\geq M \ln(1+t) \left[1 - Me^{-NK\eta}\right], \end{aligned} \quad (88)$$

where $\eta \triangleq \Pr\{\text{SINR}_{i,k}^{(m)} \leq t\} = \frac{e^{-\frac{Mt}{P}}}{(1+t)^{M-1}}$ [14]. Setting $t = \frac{P}{M}\left[\ln N - (M-1)\ln\frac{P}{M} - M \ln \ln N\right]$, it is easy to show that $\eta \geq \frac{\ln N}{N}$ and hence,

$$\mathcal{R}_{\text{RBF}} \geq M \ln\left(1 + \frac{P}{M}\left[\ln N - (M-1)\ln\frac{P}{M} - M \ln \ln N\right]\right)\left(1 - \frac{M}{N^K}\right). \quad (89)$$

Since $\ln P \nsim \Omega(\ln N)$, it follows from the above equation that $\lim_{N \to \infty} \mathcal{R}_{\text{Opt}} - \mathcal{R}_{\text{RBF}} = 0$.
∎

Theorem 9 implies that the Random Beam-forming scheme is not capable of achieving the maximum sum-rate when $\ln P \sim \Omega(\ln N)$. In other words, the Random Beam-forming scheme is not efficient in the high SNR regime. In fact, it is easy to show that the multiplexing gain of this scheme is zero. In the region of $\ln P \sim o(\ln N)$, following the approach of Theorem 3, it can be shown that with the number of feedback bits scaling as $M \lceil \log_2 M \rceil \ln \ln(P \ln N) + \omega(1)$, the maximum sum-rate capacity can be achieved.

The weak performance of Random Beam-Forming in the high SNR regime is due to the fact that the interference from the other users dominates the noise term. It can be shown that in order to achieve the maximum sum-rate, we must have $\lim_{P \to \infty} I(P) = 0$, where $I$ denotes the interference term. In other words, the interference term must be negligible compared to the noise. The Random Beam-Forming scheme can be considered as the quantization of the users' channel vectors by $M$ orthogonal code words. Since the number of code words is fixed, the quantization error, which is translated to the interference, grows with the SNR. This suggests that at high SNRs the channel of the users must be known at the BS with higher precision. This can be performed by increasing the size of the quantization code book and more efficient methods of channel quantization. Some efficient algorithms for channel quantization have been proposed in [24] [25] [26] [27].

**Theorem 10** *Consider a MIMO-BC with $N$ users ($N \to \infty$), each equipped with $K$ receive antennas, in which the base station communicates with $M$ of them with the total power constraint $P$ ($P \to \infty$). Assume that each user quantizes its channel matrix and sends the quantization index to the transmitter. Then, for any quantization method chosen by the users, any user selection strategy and any known precoding scheme chosen by the transmitter, the necessary condition to achieve $\lim_{N \to \infty} \mathcal{R}_{\mathrm{Opt}} - \mathcal{R}_{\mathrm{Opt}}^{\mathrm{Q}} = 0$, in the case of $K < M$, is having*

$$\mathbb{E}\{\mathcal{F}_Q\} \;\gtrsim\; \ln\ln(P \ln N) + \omega(1) + \sum_{i=1}^{M-K} \left[(M-i)\ln(P \ln N) - \ln N + \omega(1)\right]^+, \quad (90)$$

*and in the case of $K = M$ is having*

$$\mathbb{E}\{\mathcal{F}_Q\} \;\gtrsim\; \ln\ln\ln N + \omega(1), \quad (91)$$

*where $\mathcal{F}_{\mathrm{Q}}$ and $\mathcal{R}_{\mathrm{Opt}}^{\mathrm{Q}}$ are the total number of bits fed back to the BS, and the maximum achievable sum-rate, when the BS only has the quantized CSI, respectively, and $a^+ \triangleq \max(0, a)$.*

**Proof -** In order to prove the theorem, we assume that the BS selects $M$ users, and transmits Gaussian signals $\mathbf{x}_1, \cdots, \mathbf{x}_M$, with covariance matrices $\mathbf{Q}_1, \cdots, \mathbf{Q}_M$, respectively. Since for a fixed set of transmit covariance matrices, Dirty-Paper Coding is proved to achieve the Marton's region [5] (which is proved to be the highest known achievable region in BC), we consider this coding scheme for the proof of this theorem. In Lemmas 1-3, we state the necessary conditions for the transmit covariance matrices and the selected users, in order to achieve the maximum sum-rate capacity. Then, in Lemma 4, we associate those conditions with the size of quantization codebooks, utilized for the quantization of the selected users' channel matrices. Combining the results of the lemmas, the theorem is proved.

**Lemma 1** *The transmit covariance matrices maximizing the sum-rate capacity, in a MIMO-BC with $N \to \infty$ users, are rank one, with probability one.*

**Proof -** Assume that the selected users are indexed by $1$ to $M$. Then, the sum-rate capacity can be written as [3]

$$\mathcal{R}_{\text{Opt}} = \mathbb{E}\left\{ \max_{\substack{\mathbf{Q}_{i},\pi \\ \sum \text{Tr}\{\mathbf{Q}_i\} \leq P}} \sum_{i=1}^{M} \ln \left| \mathbf{I} + \mathbf{H}_{\pi(i)} \mathbf{Q}_{\pi(i)} \mathbf{H}_{\pi(i)}^{H} \left( \mathbf{I} + \mathbf{H}_{\pi(i)} \left( \sum_{j>i} \mathbf{Q}_{\pi(i)} \right) \mathbf{H}_{\pi(i)}^{H} \right)^{-1} \right| \right\}, \tag{92}$$

where the expectation is taken over the channel matrices $\mathbf{H}_1, \cdots, \mathbf{H}_M$. Using the duality between the MIMO-BC and MIMO Multiple Access Channel (MIMO-MAC), expressed in [3], the sum-rate capacity can be written as follows:

$$\mathcal{R}_{\text{Opt}} = \mathbb{E}_{\mathbf{H}_1,\cdots,\mathbf{H}_M} \max_{\substack{\mathbf{P}_i \\ \sum \text{Tr}\{\mathbf{P}_i\} \leq P}} \ln \left| \mathbf{I} + \sum_{i=1}^{M} \mathbf{H}_i^H \mathbf{P}_i \mathbf{H}_i \right|, \tag{93}$$

where $\mathbf{P}_i$'s are the corresponding covariance matrices in the dual MIMO-MAC. We first prove that to achieve the maximum sum-rate capacity, $\mathbf{P}_i$'s must be rank one, with probability one.

Since $\mathbf{P}_i$'s are positive semi-definite, we can write them as $\mathbf{U}_i^H \mathbf{\Lambda}_i \mathbf{U}_i$, for some unitary matrix $\mathbf{U}_i$ and diagonal matrix $\mathbf{\Lambda}_i$. Defining $\mathbf{Z}_i \triangleq \mathbf{U}_i \mathbf{H}_i$ and writing $\mathbf{\Lambda}_i = \text{Diag}(\rho_{i1}, \cdots, \rho_{iK})$, we have

$$\begin{aligned} \ln \left| \mathbf{I} + \sum_{i=1}^{M} \mathbf{H}_i^H \mathbf{P}_i \mathbf{H}_i \right| &= \ln \left| \mathbf{I} + \sum_{i=1}^{M} \mathbf{Z}_i^H \mathbf{\Lambda}_i \mathbf{Z}_i \right| \\ &= \ln \left| \mathbf{I} + \sum_{i=1}^{M} \sum_{l=1}^{K} \rho_{il} \mathbf{Z}_i(l)^H \mathbf{Z}_i(l) \right|, \end{aligned} \tag{94}$$

where $\mathbf{Z}_i(l)$ denotes the $l$th row of $\mathbf{Z}_i$. Having the fact that $|\mathbf{A}| \leq \left( \frac{\text{Tr}(\mathbf{A})}{M} \right)^M$ for any positive semi-definite matrix $\mathbf{A}$, the right hand side of the above equation can be upper-bounded as

$$\ln \left| \mathbf{I} + \sum_{i=1}^{M} \sum_{l=1}^{K} \rho_{il} \mathbf{Z}_i(l)^H \mathbf{Z}_i(l) \right| \leq M \ln \left( 1 + \frac{\sum_{i=1}^{M} \sum_{l=1}^{K} \rho_{il} \|\mathbf{Z}_i(l)\|^2}{M} \right). \tag{95}$$

Now, assume that there exists a user $k$, such that $\rho_{kl} \sim \Theta(P)$ and $\rho_{kj} \sim \Theta(P)$, for some $1 \leq l, j \leq K$. In other words, this matrix is asymptotically of rank at least 2. We have

$$\begin{aligned} \|\mathbf{Z}_k(l)\|^2 + \|\mathbf{Z}_k(j)\|^2 &\leq \|\mathbf{Z}_k\|^2 \\ &= \|\mathbf{H}_k\|^2. \end{aligned} \tag{96}$$

In [14], it has been shown that $\|\mathbf{H}_k\|_{\max}^2 < \ln N + MK \ln \ln N$, with probability one. This incurs that at least one of $\|\mathbf{Z}_k(l)\|^2$ and $\|\mathbf{Z}_k(j)\|^2$ must be less than $\frac{\ln N + MK \ln \ln N}{2}$. Without loss of generality, assume that $\|\mathbf{Z}_k(j)\|^2 < \frac{\ln N + MK \ln \ln N}{2}$. Having $\rho_{kj}$ allocated to the coordinate $(k, j)$ and using (95), yields

$$\max_{\substack{\rho_{il} \\ (i,l) \neq (k,j) \\ \sum \rho_{il} = P - \rho_{kj}}} \ln \left| \mathbf{I} + \sum_{i=1}^{M} \sum_{l=1}^{K} \rho_{il} \mathbf{Z}_i(l)^H \mathbf{Z}_i(l) \right| \leq \max_{\substack{\rho_{il} \\ (i,l) \neq (k,j) \\ \sum \rho_{il} = P - \rho_{kj}}} M \ln \left( 1 + \frac{\sum_{i=1}^{M} \sum_{l=1}^{K} \rho_{il} \|\mathbf{Z}_i(l)\|^2}{M} \right)$$

$$= M \ln \left( 1 + \frac{\max_{\substack{\rho_{il} \\ (i,l) \neq (k,j) \\ \sum \rho_{il} = P - \rho_{kj}}} \sum_{i=1}^{M} \sum_{l=1}^{K} \rho_{il} \|\mathbf{Z}_i(l)\|^2}{M} \right)$$

$$\overset{(a)}{\leq} M \ln \left( 1 + \frac{\rho_{kj}}{2M} \ln N + O(\ln \ln N) + \frac{P - \rho_{kj}}{M} \|\mathbf{Z}\|_{\max}^2 \right), \quad (97)$$

where $\|\mathbf{Z}\|_{\max}^2 \triangleq \max_{i,l} \|\mathbf{Z}_i(l)\|^2$. $(a)$ comes from the fact that the solution to the maximization problem in the second line is to allocate the rest of the available power $(P - \rho_{kj})$ to the coordinate with the highest norm. By a similar argument as before, we can show that $\|\mathbf{Z}\|_{\max}^2 < \ln N + MK \ln \ln N$, with probability one. Hence, using the above equation,

$$\text{RH (97)} \leq M \ln \left( 1 + \frac{P - \frac{\rho_{kj}}{2}}{M} [\ln N + O(\ln \ln N)] \right). \quad (98)$$

Having the fact that $\mathcal{R}_{\text{Opt}} \sim M \ln \left( \frac{P}{M} \ln N + O(\ln \ln N) \right)$, and using the above equation, we have

$$\mathcal{R}_{\text{Opt}} - \text{RH (97)} \geq M \ln \left( 1 - \frac{\rho_{kj}}{2P} \right) + O \left( \frac{\ln \ln N}{\ln N} \right). \quad (99)$$

Hence, having $\rho_{kj} \sim \Theta(P)$, incurs $\lim_{N \to \infty} \mathcal{R}_{\text{Opt}} - \text{RH (97)} > 0$. In other words, in order to have $\lim_{N \to \infty} \mathcal{R}_{\text{Opt}} - \text{RH (97)} = 0$, for each user $k$, there must be at most one $\rho_{km}$ scaling as $\Theta(P)$, and the rest must scale as $o(P)$. In the following, we will show that with probability one, for each user exactly one $\rho_{km}$ is non-zero, and the rest are zero.

Using (95) and having the fact that $\sum_{i=1}^{K} \|\mathbf{Z}_k(i)\|^2 < \ln N + MK \ln \ln N$ with probability one, it follows that the right hand side of (95) is upper-bounded by $M \ln \left( \frac{P}{M} \ln N \right)$, which is proved to be the maximum achievable sum-rate throughput in MIMO-MAC. Hence, in order to achieve the maximum sum-rate, the inequality in (95) must be turned into the equality, which means that $\sum_{i=1}^{M} \sum_{l=1}^{K} \rho_{il} \mathbf{Z}_i(l)^H \mathbf{Z}_i(l)$ must behave like $\frac{P}{M} \ln N (\mathbf{I} + o(\mathbf{I}))^5$. Moreover, since from each user at most one singular value can scale as fast as $\ln N$ [21], it follows that the maximum singular values of the selected users must scale as $\ln N$, and their corresponding powers must scale as $\frac{P}{M} + o(P)$.

---

[5]$\mathbf{A} \sim o(\mathbf{I})$ means that all the singular values of $\mathbf{A}$ are $o(1)$.

Now, assume that there exists $i, l$ such that $\lim_{N \to \infty} \frac{\|\mathbf{Z}_i(l)\|^2}{\ln N} < 1$, but $\rho_{il} \neq 0$. In the above, we have seen that $\rho_{il} \sim o(P)$. The sum-rate can be upper-bounded as

$$
\begin{aligned}
\mathcal{R} &\leq \mathcal{R}_{\text{Opt}}(P - \rho_{il}) + \ln \left| \mathbf{I} + \rho_{il} \|\mathbf{Z}_i(l)\|^2 \phi_i(l)^H \phi_i(l) \left( \mathbf{I} + \sum_{\substack{(j,m) \\ (j,m) \neq (i,l)}} \rho_{jm} \mathbf{Z}_j(m)^H \mathbf{Z}_j(m) \right)^{-1} \right| \\
&\stackrel{(a)}{\sim} \mathcal{R}_{\text{Opt}}(P - \rho_{il}) + \ln \left| \mathbf{I} + \frac{\rho_{il} \|\mathbf{Z}_i(l)\|^2}{\frac{P - \rho_{il}}{M} \ln N(1 + o(1))} \phi_i(l)^H \phi_i(l) \right| \\
&\stackrel{(b)}{\sim} M \ln \left( \frac{P - \rho_{il}}{M} \ln N(1 + o(1)) \right) + \ln \left( 1 + \frac{\rho_{il} \|\mathbf{Z}_i(l)\|^2}{\frac{P - \rho_{il}}{M} \ln N(1 + o(1))} \right) \\
&\stackrel{(c)}{\sim} M \ln \left( \frac{P}{M} \ln N(1 + o(1)) \right) - \frac{M \rho_{il}}{P} \left( 1 - \frac{\|\mathbf{Z}_i(l)\|^2}{\ln N} \right) + o(\frac{\rho_{il}}{P}),
\end{aligned}
\quad (100)
$$

where $\phi_i(l) \triangleq \frac{\mathbf{Z}_i(l)}{\|\mathbf{Z}_i(l)\|}$, and $\mathcal{R}_{\text{Opt}}(P - \rho_{il})$ denotes the maximum sum-rate when the power constraint is $P - \rho_{il}$. $(a)$ comes from the fact that achieving the maximum throughput of $\mathcal{R}_{\text{Opt}}(P - \rho_{il})$ requires that $\left( \mathbf{I} + \sum_{\substack{(j,m) \\ (j,m) \neq (i,l)}} \rho_{jm} \mathbf{Z}_j(m)^H \mathbf{Z}_j(m) \right) \sim \frac{P - \rho_{il}}{M} \ln N \left( \mathbf{I} + o(\mathbf{I}) \right)$. $(b)$ comes from the fact that $\mathcal{R}_{\text{Opt}}(P - \rho_{il}) \sim M \ln \left( \frac{P - \rho_{il}}{M} \ln N(1 + o(1)) \right)$, and finally $(c)$ results from the fact that $\rho_{il} \sim o(P)$, and using the approximation $\ln(1 + x) \approx x$, for $x \ll 1$. Suppose that instead of allocating $\rho_{il}$ to the coordinate $(i, l)$, it is allocated to the coordinate corresponding to the maximum eigenvalue of any of the selected users. Let us denote the achievable sum-rate of the system in this case by $\mathcal{R}^*$. Since the maximum singular values of the selected users scale as $\ln N$, the second term in the last line of the above equation scales as $o(\frac{\rho_{il}}{P})$ and we have

$$
\mathcal{R}^* - \mathcal{R} \sim \frac{M \rho_{il}}{P} \left( 1 - \frac{\|\mathbf{Z}_i(l)\|^2}{\ln N} \right) + o(\frac{\rho_{il}}{P}). \quad (101)
$$

As a result, if $\rho_{il} > 0$, $\mathcal{R}^* > \mathcal{R}$, which incurs that in order to achieve the maximum sum-rate $\rho_{il}$ must be zero with probability one. Having this and the fact that from each user at most one coordinate has the gain scaling as fast as $\ln N$ with probability one [21], it follows that to achieve the maximum sum-rate in the dual MIMO-MAC, the transmit covariance matrices must be rank one with probability one. Using the result of [3], the following equation holds between the covariance matrix of the user with the encoding order $j$ in the MIMO-BC, denoted by $\mathbf{Q}_{\pi(j)}$, and the covariance matrix of the user with the reverse decoding order $j$ in the dual MIMO-MAC, denoted by $\mathbf{P}_{\pi(j)}$:

$$\mathbf{Q}_{\pi(j)} = \mathbf{M}_{\pi(j)}\mathbf{P}_{\pi(j)}\mathbf{M}_{\pi(j)}^H, \tag{102}$$

where $\mathbf{M}_{\pi(j)}$ is an $M \times K$ matrix. Since $\mathbf{P}_{\pi(j)}$ is proved to be a rank one matrix with probability one, it follows from the above equation that $\mathbf{Q}_{\pi(j)}$ is also rank one with probability one, which completes the proof of Lemma 1.

∎

Lemma 1 implies that the transmit covariance matrix for the $j$th user can be written as

$$\mathbf{Q}_j = \rho_j \mathbf{\Phi}_j \mathbf{\Phi}_j^H, \tag{103}$$

where $\mathbf{\Phi}_j$ is a unit vector and $\rho_j$ is the allocated power to the $j$th user.

**Lemma 2** *The necessary condition for achieving the maximum sum-rate is that $\{\mathbf{\Phi}_j\}_{j=1}^M$, defined in the above equation, form a semi-orthogonal basis for $\mathbb{C}^M$, i.e, $|\mathbf{\Phi}_j^H \mathbf{\Phi}_i| \sim o(1)$, $i \neq j$, with probability one.*

**Proof -** The sum-rate can be upper-bounded as

$$\begin{aligned}
\mathcal{R} &\stackrel{(a)}{\leq} \mathbb{E}\left\{\sum_{i=1}^M \ln|\mathbf{I} + \mathbf{H}_i \mathbf{Q}_i \mathbf{H}_i^H|\right\} \\
&\stackrel{(103)}{=} \mathbb{E}\left\{\sum_{i=1}^M \ln|\mathbf{I} + \rho_i \mathbf{H}_i \mathbf{\Phi}_i \mathbf{\Phi}_i^H \mathbf{H}_i^H|\right\} \\
&= \mathbb{E}\left\{\sum_{i=1}^M \ln\left(1 + \rho_i \|\mathbf{H}_i \mathbf{\Phi}_i\|^2\right)\right\} \\
&= \mathbb{E}\left\{\sum_{i=1}^M \ln\left(1 + \rho_i \sum_{l=1}^K \lambda_l(i) \left|\mathbf{v}_l^H(i)\mathbf{\Phi}_i\right|^2\right)\right\} \\
&= \mathbb{E}\left\{\sum_{i=1}^M \ln\left(1 + \rho_i \left[\lambda_1(i)\left|\mathbf{v}_1^H(i)\mathbf{\Phi}_i\right|^2 + \sum_{l=2}^K \lambda_l(i)\left|\mathbf{v}_l^H(i)\mathbf{\Phi}_i\right|^2\right]\right)\right\}, \tag{104}
\end{aligned}$$

where $(a)$ comes from ignoring the interference terms, $\lambda_l(i)$ denotes the $l$th ordered singular value of $\mathbf{H}_i \mathbf{H}_i^H$, and $\mathbf{v}_l(i)$ denotes its corresponding eigenvector. Having the facts that $\lambda_1(i) \sim \ln N + o(\ln N)$, which has been proved to be the necessary condition to achieve the maximum sum-rate (in Lemma 1), and $\|\mathbf{H}_i\|^2 = \sum_l \lambda_l(i) \sim \ln N + o(\ln N)$, with probability one [14], it follows that $\sum_{l=2}^K \lambda_l(i)\left|\mathbf{v}_l^H(i)\mathbf{\Phi}_i\right|^2 \sim o(\ln N)$. Having this and $\mathcal{R}_{\text{Opt}} \sim M \ln\left(\frac{P}{M}\ln N + o(\ln N)\right)$ [14], it

follows that to achieve the maximum sum-rate we must have $\lambda_1(i) \left|\mathbf{v}_1^H(i)\mathbf{\Phi}_i\right|^2 \sim \ln N[1+o(1)]$, $\forall i, 1 \leq i \leq M$. Noting $\lambda_1(i) \sim \ln N + O(\ln \ln N)$, we conclude $\left|\mathbf{v}_1^H(i)\mathbf{\Phi}_i\right|^2 \sim 1 + o(1)$, $\forall 1 \leq i \leq M$. In other words, the coordinate of the transmit covariance matrix for each user is almost in the direction of the eigenvector corresponding to the maximum singular value of that user.

The rate of the $i$th encoded user can be upper-bounded as

$$\begin{aligned}
\mathcal{R}_{\pi(i)} &= \mathbb{E}\left\{\ln\left|\mathbf{I} + \mathbf{H}_{\pi(i)}\mathbf{Q}_{\pi(i)}\mathbf{H}_{\pi(i)}^H\left(\mathbf{I} + \mathbf{H}_{\pi(i)}\left(\sum_{j>i}\mathbf{Q}_{\pi(j)}\right)\mathbf{H}_{\pi(i)}^H\right)^{-1}\right|\right\} \\
&\leq \frac{1}{M-i}\sum_{j=i+1}^M \mathbb{E}\left\{\ln\left|\mathbf{I} + \mathbf{H}_{\pi(i)}\mathbf{Q}_{\pi(i)}\mathbf{H}_{\pi(i)}^H\left(\mathbf{I} + \mathbf{H}_{\pi(i)}\mathbf{Q}_{\pi(j)}\mathbf{H}_{\pi(i)}^H\right)^{-1}\right|\right\}.
\end{aligned} \quad (105)$$

Substituting $\mathbf{Q}_{\pi(i)}$ and $\mathbf{Q}_{\pi(j)}$ from (103) yields

$$\begin{aligned}
\mathcal{R}_{\pi(i)} &\leq \frac{1}{M-i}\sum_{j=i+1}^M \mathbb{E}\left\{\ln\left|\mathbf{I} + \rho_{\pi(i)}\eta_{\pi(i)}\mathbf{\Psi}_{\pi(i)}\mathbf{\Psi}_{\pi(i)}^H\left(\mathbf{I} + \rho_{\pi(j)}I_{\pi(j)}^{\pi(i)}\mathbf{\Omega}_{\pi(j)}\mathbf{\Omega}_{\pi(j)}^H\right)^{-1}\right|\right\} \\
&\stackrel{(a)}{=} \frac{1}{M-i}\sum_{j=i+1}^M \mathbb{E}\left\{\ln\left(1 + \rho_{\pi(i)}\eta_{\pi(i)}\mathbf{\Psi}_{\pi(i)}^H\left[\mathbf{I} - \frac{\rho_{\pi(j)}I_{\pi(j)}^{\pi(i)}}{1+\rho_{\pi(j)}I_{\pi(j)}^{\pi(i)}}\mathbf{\Omega}_{\pi(j)}\mathbf{\Omega}_{\pi(j)}^H\right]\mathbf{\Psi}_{\pi(i)}\right)\right\} \\
&= \frac{1}{M-i}\sum_{j=i+1}^M \mathbb{E}\left\{\ln\left(1 + \rho_{\pi(i)}\eta_{\pi(i)}\right)\right\} + \\
&\quad \mathbb{E}\left\{\ln\left(1 - \frac{\rho_{\pi(i)}\eta_{\pi(i)}}{1+\rho_{\pi(i)}\eta_{\pi(i)}}\frac{\rho_{\pi(j)}I_{\pi(j)}^{\pi(i)}}{1+\rho_{\pi(j)}I_{\pi(j)}^{\pi(i)}}|\mathbf{\Psi}_{\pi(i)}^H\mathbf{\Omega}_{\pi(j)}|^2\right)\right\} \\
&\stackrel{(b)}{\leq} \frac{1}{M-i}\sum_{j=i+1}^M \mathbb{E}\left\{\ln\left(1 + \rho_{\pi(i)}\eta_{\pi(i)}\right)\right\} + \\
&\quad \ln\left(1 - \mathbb{E}\left\{\frac{\rho_{\pi(i)}\eta_{\pi(i)}}{1+\rho_{\pi(i)}\eta_{\pi(i)}}\frac{\rho_{\pi(j)}I_{\pi(j)}^{\pi(i)}}{1+\rho_{\pi(j)}I_{\pi(j)}^{\pi(i)}}|\mathbf{\Psi}_{\pi(i)}^H\mathbf{\Omega}_{\pi(j)}|^2\right\}\right),
\end{aligned} \quad (106)$$

where $\eta_{\pi(i)} \triangleq \|\mathbf{H}_{\pi(i)}\mathbf{\Phi}_{\pi(i)}\|^2$, $I_{\pi(j)}^{\pi(i)} \triangleq \|\mathbf{H}_{\pi(i)}\mathbf{\Phi}_{\pi(j)}\|^2$, $\mathbf{\Psi}_{\pi(i)} \triangleq \frac{\mathbf{H}_{\pi(i)}\mathbf{\Phi}_{\pi(i)}}{\|\mathbf{H}_{\pi(i)}\mathbf{\Phi}_{\pi(i)}\|}$, $\mathbf{\Omega}_{\pi(j)} \triangleq \frac{\mathbf{H}_{\pi(i)}\mathbf{\Phi}_{\pi(j)}}{\|\mathbf{H}_{\pi(i)}\mathbf{\Phi}_{\pi(j)}\|}$. $(a)$ comes from the facts $|\mathbf{I} + \mathbf{AB}| = |\mathbf{I} + \mathbf{BA}|$ and $\left(\mathbf{I} + \rho_{\pi(j)}I_{\pi(j)}^{\pi(i)}\mathbf{\Omega}_{\pi(j)}\mathbf{\Omega}_{\pi(j)}^H\right)^{-1} = \mathbf{I} - \frac{\rho_{\pi(j)}I_{\pi(j)}^{\pi(i)}}{1+\rho_{\pi(j)}I_{\pi(j)}^{\pi(i)}}\mathbf{\Omega}_{\pi(j)}\mathbf{\Omega}_{\pi(j)}^H$, and $(b)$ comes from the concavity of $\ln$ function. From the above equation, and noting the facts that $\mathbb{E}\left\{\ln\left(1+\rho_{\pi(i)}\eta_{\pi(i)}\right)\right\} \lesssim \ln\left(\frac{P}{M}\ln N + o(\ln N)\right)$ and $\mathcal{R}_{\text{Opt}} \sim M\ln\left(\frac{P}{M}\ln N + o(\ln N)\right)$, it follows that in order to achieve the maximum sum-rate, the term

$$\ln\left(1 - \mathbb{E}\left\{\frac{\rho_{\pi(i)}\eta_{\pi(i)}}{1+\rho_{\pi(i)}\eta_{\pi(i)}}\frac{\rho_{\pi(j)}I_{\pi(j)}^{\pi(i)}}{1+\rho_{\pi(j)}I_{\pi(j)}^{\pi(i)}}|\mathbf{\Psi}_{\pi(i)}^H\mathbf{\Omega}_{\pi(j)}|^2\right\}\right)$$

must approach zero for all $i$ and $j > i$, which incurs that $\mathbb{E}\left\{\frac{\rho_{\pi(i)}\eta_{\pi(i)}}{1+\rho_{\pi(i)}\eta_{\pi(i)}}\frac{\rho_{\pi(j)}I_{\pi(j)}^{\pi(i)}}{1+\rho_{\pi(j)}I_{\pi(j)}^{\pi(i)}}|\boldsymbol{\Psi}_{\pi(i)}^H\boldsymbol{\Omega}_{\pi(j)}|^2\right\} \sim o(1), \forall i, j > i$. Since $\rho_{\pi(i)} \to \infty$ (as $P \to \infty$), and $\eta_{\pi(i)} \sim \ln N$, the term $\frac{\rho_{\pi(i)}\eta_{\pi(i)}}{1+\rho_{\pi(i)}\eta_{\pi(i)}} \approx 1$, with probability one. Writing $\mathbf{v}_1(\pi(i))$ as $\alpha_{\pi(i)}\boldsymbol{\Phi}_{\pi(i)} + \mathbf{v}_1(\pi(i))^\perp$ and $\boldsymbol{\Phi}_{\pi(i)}$ as $\gamma_{\pi(i)}\mathbf{v}_1(\pi(i)) + \boldsymbol{\Phi}_{\pi(i)}^\perp$, where $\alpha_{\pi(i)} \triangleq \boldsymbol{\Phi}_{\pi(i)}^H\mathbf{v}_1(\pi(i))$, $\gamma_{\pi(i)} \triangleq \mathbf{v}_1(\pi(i))^H\boldsymbol{\Phi}_{\pi(i)}$, $\mathbf{v}_1(\pi(i))^\perp$ denotes the projection of $\mathbf{v}_1(\pi(i))$ over the null space of $\boldsymbol{\Phi}_{\pi(i)}$ and $\boldsymbol{\Phi}_{\pi(i)}^\perp$ denotes the projection of $\boldsymbol{\Phi}_{\pi(i)}$ over the null space of $\mathbf{v}_1(\pi(i))$, $\chi \triangleq \mathbb{E}\left\{\frac{\rho_{\pi(j)}I_{\pi(j)}^{\pi(i)}}{1+\rho_{\pi(j)}I_{\pi(j)}^{\pi(i)}}|\boldsymbol{\Psi}_{\pi(i)}^H\boldsymbol{\Omega}_{\pi(j)}|^2\right\}$ can be written as

$$\begin{aligned}
\chi &= \mathbb{E}\left\{\frac{\rho_{\pi(j)}I_{\pi(j)}^{\pi(i)}}{1+\rho_{\pi(j)}I_{\pi(j)}^{\pi(i)}}\frac{\left|\boldsymbol{\Phi}_{\pi(i)}^H\mathbf{H}_{\pi(i)}^H\mathbf{H}_{\pi(i)}\boldsymbol{\Phi}_{\pi(j)}\right|^2}{\eta_{\pi(i)}I_{\pi(j)}^{\pi(i)}}\right\} \\
&= \mathbb{E}\left\{\frac{\rho_{\pi(j)}}{1+\rho_{\pi(j)}I_{\pi(j)}^{\pi(i)}}\frac{\left|\left[\gamma_{\pi(i)}\mathbf{v}_1(\pi(i)) + \boldsymbol{\Phi}_{\pi(i)}^\perp\right]^H\mathbf{H}_{\pi(i)}^H\mathbf{H}_{\pi(i)}\boldsymbol{\Phi}_{\pi(j)}\right|^2}{\eta_{\pi(i)}}\right\} \\
&\overset{(a)}{\geq} \mathbb{E}\left\{\frac{\rho_{\pi(j)}}{1+\rho_{\pi(j)}I_{\pi(j)}^{\pi(i)}}\frac{\left(|\gamma_{\pi(i)}|\left|\mathbf{v}_1(\pi(i))^H\mathbf{H}_{\pi(i)}^H\mathbf{H}_{\pi(i)}\boldsymbol{\Phi}_{\pi(j)}\right| - \left|\left(\boldsymbol{\Phi}_{\pi(i)}^\perp\right)^H\mathbf{H}_{\pi(i)}^H\mathbf{H}_{\pi(i)}\boldsymbol{\Phi}_{\pi(j)}\right|\right)^2}{\eta_{\pi(i)}}\right\} \\
&\overset{(b)}{\geq} \mathbb{E}\left\{\frac{\rho_{\pi(j)}}{1+\rho_{\pi(j)}I_{\pi(j)}^{\pi(i)}}\frac{\left(|\gamma_{\pi(i)}|\lambda_{\max}(\pi(i))\left|\mathbf{v}_1(\pi(i))^H\boldsymbol{\Phi}_{\pi(j)}\right| - \lambda_{\max}(\pi(i))\|\boldsymbol{\Phi}_{\pi(i)}^\perp\|\right)^2}{\eta_{\pi(i)}}\right\} \\
&\overset{(c)}{\geq} \mathbb{E}\left\{\frac{\rho_{\pi(j)}\lambda_{\max}(\pi(i))}{1+\rho_{\pi(j)}I_{\pi(j)}^{\pi(i)}}\left(|\gamma_{\pi(i)}|\left|\left[\alpha_{\pi(i)}\boldsymbol{\Phi}_{\pi(i)} + \mathbf{v}_1(\pi(i))^\perp\right]^H\boldsymbol{\Phi}_{\pi(j)}\right| - \|\boldsymbol{\Phi}_{\pi(i)}^\perp\|\right)^2\right\} \\
&\overset{(d)}{\geq} \mathbb{E}\left\{\left(|\gamma_{\pi(i)}||\alpha_{\pi(i)}|\left|\boldsymbol{\Phi}_{\pi(i)}^H\boldsymbol{\Phi}_{\pi(j)}\right| - \|\mathbf{v}_1(\pi(i))^\perp\| - \|\boldsymbol{\Phi}_{\pi(i)}^\perp\|\right)^2\right\}, \quad (107)
\end{aligned}$$

where $\lambda_{\max}(\pi(i))$ denotes the maximum singular value of $\mathbf{H}_{\pi(i)}^H\mathbf{H}_{\pi(i)}$. $(a)$ comes from the fact that $|a+b|^2 \geq (|a|-|b|)^2$. $(b)$ results from the facts that $\mathbf{v}_1(\pi(i))$ is the eigenvector corresponding to the maximum singular value of $\mathbf{H}_{\pi(i)}$, and hence, $\mathbf{v}_1(\pi(i))^H\mathbf{H}_{\pi(i)}^H\mathbf{H}_{\pi(i)} = \lambda_{\max}(\pi(i))\mathbf{v}_1(\pi(i))^H$, and also $\left|\left(\boldsymbol{\Phi}_{\pi(i)}^\perp\right)^H\mathbf{H}_{\pi(i)}^H\mathbf{H}_{\pi(i)}\boldsymbol{\Phi}_{\pi(j)}\right|^2 \leq \|\boldsymbol{\Phi}_{\pi(i)}^\perp\|^2\lambda_{\max}(\pi(i))$. $(c)$ comes from the fact that $\eta_{\pi(i)} = \|\mathbf{H}_{\pi(i)}\boldsymbol{\Phi}_{\pi(i)}\|^2 \leq \lambda_{\max}(\pi(i))$, and finally $(d)$ results from the facts that $I_{\pi(j)}^{\pi(i)} = \|\mathbf{H}_{\pi(i)}\boldsymbol{\Phi}_{\pi(j)}\|^2 \leq \lambda_{\max}(\pi(i))$, $\left|\left[\alpha_{\pi(i)}\boldsymbol{\Phi}_{\pi(i)} + \mathbf{v}_1(\pi(i))^\perp\right]^H\boldsymbol{\Phi}_{\pi(j)}\right| \geq |\alpha_{\pi(i)}|\left|\boldsymbol{\Phi}_{\pi(i)}^H\boldsymbol{\Phi}_{\pi(j)}\right| - \|\mathbf{v}_1(\pi(i))^\perp\|$, and $|\gamma_{\pi(i)}| < 1$. Since $\left|\mathbf{v}_1^H(\pi(i))\boldsymbol{\Phi}_{\pi(i)}\right| \sim 1+o(1)$, it follows that $|\alpha_{\pi(i)}| = |\gamma_{\pi(i)}| \sim 1+o(1)$ and $\|\mathbf{v}_1(\pi(i))^\perp\| = \left\|\boldsymbol{\Phi}_{\pi(i)}^\perp\right\| \sim o(1)$. Hence, the necessary condition to achieve the maximum sum-rate is having

$\left|\mathbf{\Phi}_{\pi(i)}^H\mathbf{\Phi}_{\pi(j)}\right|^2 \sim o(1)$, $\forall i, j > i$, with probability one. In other words, $\mathbf{\Phi}_{\pi(i)}$ and $\mathbf{\Phi}_{\pi(j)}$ must be semi-orthogonal to each other with probability one, which completes the proof of Lemma 2.

∎

*Remark* - It is worth to note that the right hand side of (104) achieves the maximum sum-rate of $M \ln\left(1 + \frac{P}{M} \ln N[1 + o(1)]\right)$ if the power is uniformly allocated to the coordinates, almost surely. In other words, $\rho_i = \frac{P}{M}[1 + o(1)]$.

**Lemma 3** *Defining $\epsilon_i \triangleq \mathbf{v}_1^H(\pi(i))\mathbf{\Upsilon}_i$, where $\mathbf{\Upsilon}_i \triangleq \left[\mathbf{\Phi}_{\pi(i+1)}|\cdots|\mathbf{\Phi}_{\pi(M)}\right]$, and $\mathbf{v}_1(\pi(i))$ denotes the eigenvector corresponding to the maximum eigenvalue of the $i$th encoded user, assuming Dirty-paper Coding, the necessary condition to have $\mathcal{R}_{\mathrm{Opt}} - \mathcal{R} \to 0$, in the case $K < M - i + 1$ is $\|\epsilon_i\|^2 \sim o\left(\frac{1}{P \ln N}\right)$ and in the case $K \geq M - i + 1$ is $\|\epsilon_i\|^2 \sim o(1)$, with probability one.*

**Proof** - Consider the user with the encoding order $i$. The rate of this user can be upper-bounded as

$$\begin{aligned}
\mathcal{R}_{\pi(i)} &\leq \mathbb{E}\left\{\ln\left|\mathbf{I} + \mathbf{H}_{\pi(i)}\mathbf{Q}_{\pi(i)}\mathbf{H}_{\pi(i)}^H\left(\mathbf{I} + \mathbf{H}_{\pi(i)}\left[\sum_{j=i+1}^{M}\mathbf{Q}_{\pi(j)}\right]\mathbf{H}_{\pi(i)}^H\right)^{-1}\right|\right\} \\
&= \mathbb{E}\left\{\ln\left|\mathbf{I} + \rho_{\pi(i)}\mathbf{H}_{\pi(i)}\mathbf{\Phi}_{\pi(i)}\mathbf{\Phi}_{\pi(i)}^H\mathbf{H}_{\pi(i)}^H\left(\mathbf{I} + \mathbf{H}_{\pi(i)}\left[\sum_{j=i+1}^{M}\rho_{\pi(j)}\mathbf{\Phi}_{\pi(j)}\mathbf{\Phi}_{\pi(j)}^H\right]\mathbf{H}_{\pi(i)}^H\right)^{-1}\right|\right\}.
\end{aligned}$$
(108)

Writing the SVD of $\mathbf{H}_{\pi(i)}$ as $\mathbf{U}_{\pi(i)}\mathbf{\Lambda}_{\pi(i)}\mathbf{V}_{\pi(i)}^H$, we have

$$\mathcal{R}_{\pi(i)} \leq \mathbb{E}\left\{\ln\left|\mathbf{I} + \rho_{\pi(i)}\lambda_1(\pi(i))\mathbf{\Psi}_{\pi(i)}\mathbf{\Psi}_{\pi(i)}^H\mathbf{W}\right|\right\}, \tag{109}$$

where $\mathbf{W} \triangleq (\mathbf{I} + \mathbf{G})^{-1}$, in which $\mathbf{G} \triangleq \mathbf{\Lambda}_{\pi(i)}\mathbf{V}_{\pi(i)}^H\left[\sum_{j=i+1}^{M}\rho_{\pi(j)}\mathbf{\Phi}_{\pi(j)}\mathbf{\Phi}_{\pi(j)}^H\right]\mathbf{V}_{\pi(i)}\mathbf{\Lambda}_{\pi(i)}^T$, and $\mathbf{\Psi}_{\pi(i)} \triangleq \frac{\mathbf{\Lambda}_{\pi(i)}\mathbf{V}_{\pi(i)}^H\mathbf{\Phi}_{\pi(i)}}{\sqrt{\lambda_1(\pi(i))}}$. Having the facts that $\mathbf{v}_1^H(\pi(i))\mathbf{\Phi}_{\pi(i)} \sim 1 + o(1)$, $\mathbf{v}_j^H(\pi(i))\mathbf{\Phi}_{\pi(i)} \sim o(1)$, $j \neq 1$ (Lemma 2), $\lambda_1(\pi(i)) \sim \ln N$, and $\lambda_j(\pi(i)) \sim o(\ln N)$, $j \neq 1$ (Lemma 1), we have $\mathbf{\Psi}_{\pi(i)} = [1 + o(1), o(1), \cdots, o(1)]^T$. In other words, as $N \to \infty$, $\mathbf{\Psi}_{\pi(i)}$ approaches to the vector $[1, 0, \cdots, 0]^T$. Using $|\mathbf{I} + \mathbf{AB}| = |\mathbf{I} + \mathbf{BA}|$, we can write

$$\begin{aligned}
\mathcal{R}_{\pi(i)} &\leq \mathbb{E}\left\{\ln\left(1 + \rho_{\pi(i)}\lambda_1(\pi(i))\mathbf{\Psi}_{\pi(i)}^H\mathbf{W}\mathbf{\Psi}_{\pi(i)}\right)\right\} \\
&\approx \mathbb{E}\left\{\ln\left(1 + \rho_{\pi(i)}\lambda_1(\pi(i))\mathbf{W}_{11}[1 + o(1)]\right)\right\},
\end{aligned} \tag{110}$$

where $\mathbf{A}_{ij}$ denotes the $(i, j)$th entry of matrix $\mathbf{A}$. Using the concavity of $\ln$ function, and having the facts that $\lambda_1(\pi(i)) \sim \ln N + o(\ln N)$ with probability one, we have

$$\mathcal{R}_{\pi(i)} \leq \ln\left(1 + \rho_{\pi(i)}(\ln N)\mathbb{E}\left\{\mathbf{W}_{11}\right\}[1 + o(1)]\right). \tag{111}$$

Since the necessary condition to achieve the maximum sum-rate is having $\mathcal{R}_{\pi(i)} \sim \ln(\frac{P}{M} \ln N)$, $\forall i$, the above equation implies that the necessary condition to have $\lim_{N\to\infty} \mathcal{R}_{\text{Opt}} - \mathcal{R} = 0$ is having $\mathbb{E}\{\mathbf{W}_{11}\} \sim 1 + o(1)$, which incurs that $\mathbf{W}_{11}$ must scale as $1 + o(1)$, with probability one. In the following, we calculate $\mathbf{W}_{11}$.

$\mathbf{G} = \mathbf{\Lambda}_{\pi(i)} \mathbf{V}_{\pi(i)}^H \left[ \sum_{j=i+1}^{M} \rho_{\pi(j)} \mathbf{\Phi}_{\pi(j)} \mathbf{\Phi}_{\pi(j)}^H \right] \mathbf{V}_{\pi(i)} \mathbf{\Lambda}_{\pi(i)}^T$ can be written as

$$\mathbf{G} = \mathbf{Z}\mathbf{\Theta}\mathbf{\Theta}^H\mathbf{Z}^H, \tag{112}$$

where $\mathbf{Z} \triangleq \left[ \sqrt{\lambda_1(\pi(i))} \mathbf{v}_1(\pi(i)) | \cdots | \sqrt{\lambda_K(\pi(i))} \mathbf{v}_K(\pi(i)) \right]^H$, and

$$\mathbf{\Theta} \triangleq \left[ \sqrt{\rho_{\pi(i+1)}} \mathbf{\Phi}_{\pi(i+1)} | \cdots | \sqrt{\rho_{\pi(M)}} \mathbf{\Phi}_{\pi(M)} \right].$$

$\mathbf{Z}\mathbf{\Theta}$ can be written as $\left[ \mathbf{\Xi}^T | \mathbf{\Omega}^T \right]^T$, where $\mathbf{\Xi} \triangleq \sqrt{\lambda_1(\pi(i))} \mathbf{v}_1^H(\pi(i))\mathbf{\Theta}$ and $\mathbf{\Omega} \triangleq \mathbf{Z}_r \mathbf{\Theta}$, and $\mathbf{Z}_r \triangleq \left[ \sqrt{\lambda_2(\pi(i))} \mathbf{v}_2(\pi(i)) | \cdots | \sqrt{\lambda_K(\pi(i))} \mathbf{v}_K(\pi(i)) \right]^H$. Substituting in the above equation yields

$$\mathbf{G} = \begin{bmatrix} \|\mathbf{\Xi}\|^2 & \mathbf{\Xi}\mathbf{\Omega}^H \\ \mathbf{\Omega}\mathbf{\Xi}^H & \mathbf{\Omega}\mathbf{\Omega}^H \end{bmatrix}. \tag{113}$$

As a result, $\mathbf{W}_{11}$ can be written as

$$\begin{aligned} \mathbf{W}_{11} &= \frac{|\mathbf{I} + \mathbf{\Omega}\mathbf{\Omega}^H|}{|\mathbf{I} + \mathbf{G}|} \\ &= \frac{|\mathbf{I} + \mathbf{\Omega}\mathbf{\Omega}^H|}{(1 + \|\mathbf{\Xi}\|^2)|\mathbf{I} + \mathbf{\Omega}\mathbf{\Omega}^H| + \sum_{j=2}^{K}(-1)^{j+1}\mathbf{G}_{1j}|\Delta(\mathbf{C}_{1j})|}, \end{aligned} \tag{114}$$

where $\Delta(\mathbf{C}_{1j})$ denotes the minor of $\mathbf{C}_{1j}$ and $\mathbf{C} \triangleq \mathbf{G} + \mathbf{I}$. $|\Delta(\mathbf{C}_{1j})|$ can be computed as

$$|\Delta(\mathbf{C}_{1j})| = \sum_{\substack{i \\ 1,j \notin \mathcal{A}_i}} |\Delta_{\mathcal{A}_i}(\mathbf{G}_{1j})|, \tag{115}$$

where $\Delta_{\mathcal{A}_i}(\mathbf{G}_{1j})$ denotes a sub-matrix of $\Delta(\mathbf{G}_{1j})$, resulted from deleting the rows and columns corresponding to the elements in $\mathcal{A}_i$, and $\mathcal{A}_i$ is an arbitrary subset of $\{1, 2, \cdots, K\}$. Note that $\Delta_\emptyset(\mathbf{G}_{1j}) = \Delta(\mathbf{G}_{1j})$, where $\emptyset$ denotes the null set. Similarly, we can write

$$|\mathbf{I} + \mathbf{\Omega}\mathbf{\Omega}^H| = \sum_{\substack{i \\ 1 \notin \mathcal{A}_i}} |\Delta_{\mathcal{A}_i}(\mathbf{G}_{11})|. \tag{116}$$

Substituting (115) and (116) in (114), after some manipulations, we obtain

$$\mathbf{W}_{11} = \frac{|\mathbf{I} + \mathbf{\Omega}\mathbf{\Omega}^H|}{|\mathbf{I} + \mathbf{\Omega}\mathbf{\Omega}^H| + |\mathbf{G}| + \|\mathbf{\Xi}\|^2 \delta_1 + \sum_{j=2}^{K}(-1)^{j+1}\mathbf{G}_{1j}\delta_j}, \tag{117}$$

where $\delta_1 \triangleq \sum_{\substack{i \\ 1 \notin \mathcal{A}_i \\ \mathcal{A}_i \neq \emptyset}} |\Delta_{\mathcal{A}_i}(\mathbf{G}_{11})|$ and $\delta_j \triangleq \sum_{\substack{i \\ 1,j \notin \mathcal{A}_i \\ \mathcal{A}_i \neq \emptyset}} |\Delta_{\mathcal{A}_i}(\mathbf{G}_{1j})|$. Two situations can occur here:

- *Case I; $K \geq M - i + 1$*: In this case, since $\mathbf{G}$ is of rank at most $M - i$, $|\mathbf{G}| = 0$ in the above equation. We have observed that in order to achieve the maximum sum-rate $\rho_{\pi(j)} \sim \frac{P}{M}[1 + o(1)]$, which incurs $|\mathbf{G}_{lk}| \sim \Theta\left(Pf^{(1)}(\boldsymbol{\lambda})\right)$, $k, l \neq 1$, where $\boldsymbol{\lambda} \triangleq [\lambda_2(\pi(i)), \cdots, \lambda_K(\pi(i))]$, and $f^{(m)}(\boldsymbol{\lambda})$ denotes a function of $\boldsymbol{\lambda}$, with order $m$ [6]. Having this, it can be easily proved that

$$\|\boldsymbol{\Xi}\|^2 \delta_1 + \sum_{j=2}^{K}(-1)^{j+1}\mathbf{G}_{1j}\delta_j \sim \Theta\left(P^{K-2}\|\boldsymbol{\Xi}\|^2 f^{(K-2)}(\boldsymbol{\lambda})\right),$$

and

$$|\mathbf{I} + \boldsymbol{\Omega}\boldsymbol{\Omega}^H| \sim \Theta\left(P^{K-1} f^{(K-2)}(\boldsymbol{\lambda}) g^{(1)}(\boldsymbol{\lambda})\right). \tag{118}$$

Using this and (117), it follows that the necessary condition to satisfy $\mathbf{W}_{11} \sim 1 + o(1)$ is having $\|\boldsymbol{\Xi}\|^2 \sim o\left(Pg^{(1)}(\boldsymbol{\lambda})\right)$. Since $g^{(1)}(\boldsymbol{\lambda}) \sim o(\ln N)$, this condition can be written as $\|\boldsymbol{\Xi}\|^2 \sim o\left(P \ln N\right)$.

- *Case II; $K < M - i + 1$*: In this case, $\mathbf{G}$ is full-rank with probability one and with a similar argument as in the previous part, we can show that

$$|\mathbf{G}| \sim \Theta\left(\|\boldsymbol{\Xi}\|^2 P^{K-1} f^{(K-2)}(\boldsymbol{\lambda}) g^{(1)}(\boldsymbol{\lambda})\right).$$

Hence, using (117) and (118), the necessary condition to satisfy $\mathbf{W}_{11} \sim 1 + o(1)$ is having $\|\boldsymbol{\Xi}\|^2 \sim o\left(1\right)$.

Having the facts that $\rho_{\pi(j)} \sim \frac{P}{M}$ and $\lambda_1(\pi(i)) \sim \ln N$, we have $\|\epsilon_i\|^2 \sim \frac{\|\boldsymbol{\Xi}_i\|^2}{P \ln N}$. Therefore, the conditions of $\|\boldsymbol{\Xi}\|^2 \sim o\left(P \ln N\right)$ and $\|\boldsymbol{\Xi}\|^2 \sim o\left(1\right)$ are translated into $\|\epsilon_i\|^2 \sim o\left(1\right)$ and $\|\epsilon_i\|^2 \sim o\left(\frac{1}{P \ln N}\right)$, respectively, which completes the proof of Lemma 3.

∎

*Remark* - Note that since

$$\|\epsilon_i\|^2 = \sum_{j=i+1}^{M} |\mathbf{v}_1^H(\pi(i))\boldsymbol{\Phi}_{\pi(j)}|^2,$$

it follows that for case 1,

$$|\mathbf{v}_1^H(\pi(i))\boldsymbol{\Phi}_{\pi(j)}|^2 \sim o\left(1\right), \quad i+1 \leq j \leq M,$$

and for case 2,

$$|\mathbf{v}_1^H(\pi(i))\boldsymbol{\Phi}_{\pi(j)}|^2 \sim o\left(\frac{1}{P \ln N}\right), \quad i+1 \leq j \leq M.$$

---

[6] A function $f(x_1, \cdots, x_n)$ is said to be of order $m$, if it can be written as $\sum_j c_j \prod_{l=1}^{n} x_l^{\alpha_l(j)}$, where $\sum_{l=1}^{n} \alpha_l(j) = m$, $\forall j$.

In other words, achieving the maximum sum-rate imposes an orthogonality constraint between the eigenvector corresponding to the maximum singular value of each user and the coordinates of the transmitted signal for users with higher encoding orders. This orthogonality constraint is much more restrictive in the second case.

In Lemmas 1-3, we have proved that, for any user selection strategy and any known precoding scheme, in order to achieve the maximum sum-rate capacity, the following constraints must be satisfied with probability one:

- The maximum singular values of selected users must behave as $\ln N$.
- The transmit covariance matrices must be rank one.
- The transmit coordinates must be almost orthogonal to each other. Moreover, they must be almost in the direction of the eigenvectors corresponding to the maximum singular values of the selected users.
- The transmit power must be allocated almost uniformly among the selected users.

Having the above constraints satisfied, depending on the number of receive antennas, an orthogonality constraint must be satisfied between the eigenvector corresponding to the maximum singular value of each user and the transmit coordinates of the users with higher encoding orders, with probability one. Now, the question is that, taking the effect of quantization into account, how accurate should the BS know the channels of the selected users such that the above constraints are satisfied. For this purpose, we focus on the last constraint and associate $\|\epsilon_i\|^2$ with the size of the quantization cookbook for the $i$th encoded user in the following lemma:

**Lemma 4** *Let $L_i$ be the size of the codebook used for the quantization of $\mathbf{H}_{\pi(i)}$. Then, for any quantization method and any value of $\theta$, we have*

$$Pr\{\|\epsilon_i\|^2 > \theta\} \geq \left[\max\left(0, 1 - L_i \binom{M-1}{i-1}\theta^{M-i}\right)\right]^N. \quad (119)$$

**Proof -** Since the transmitter only knows the quantized information about the channel matrices, we can write $\mathbf{v}_1(\pi(i))$ as $\widehat{\mathbf{v}}_1(\pi(i)) + \Delta\mathbf{v}_1(\pi(i))$, where $\widehat{\mathbf{v}}_1(\pi(i))$ is perfectly known by the transmitter and can be considered as a deterministic vector, and $\Delta\mathbf{v}_1(\pi(i))$ is unknown to the transmitter. Hence, we have

$$\begin{aligned}\epsilon_i &= [\widehat{\mathbf{v}}_1(\pi(i)) + \Delta\mathbf{v}_1(\pi(i))]^H \mathbf{\Upsilon}_i \\ &= \mathbf{b}_{\pi(i)} + \Delta\mathbf{x}_{\pi(i)}, \quad (120)\end{aligned}$$

34where $\mathbf{b}_{\pi(i)} \triangleq \widehat{\mathbf{v}}_1^H(\pi(i))\mathbf{\Upsilon}_i$ is a $1 \times (M - i)$ vector, known to the transmitter, while $\Delta \mathbf{x}_{\pi(i)} \triangleq \Delta \mathbf{v}_1^H(\pi(i))\mathbf{\Upsilon}_i$ is an unknown $1 \times (M - i)$ vector. We can write

$$\|\epsilon_i\|^2 \geq \min_n \|\mathbf{b}_n + \Delta\mathbf{v}_1^H(n)\mathbf{\Upsilon}_i\|^2, \tag{121}$$

where $\mathbf{v}_1(n)$ denotes the eigenvector corresponding to the maximum singular value of the $n$th user, $\Delta \mathbf{v}_1(n)$ denotes the error in $\mathbf{v}_1(n)$ due to the quantization of $\mathbf{H}_n$, and $\mathbf{b}_n \triangleq \widehat{\mathbf{v}}_1^H(n)\mathbf{\Upsilon}_i$. In fact, in the above equation, it is assumed that all users quantize their channel matrices, and $\|\epsilon_i\|^2$ is lower-bounded by the minimum error. Since $\Delta \mathbf{v}_1(n)$ are i.i.d random variables, it follows that $\mu_n \triangleq \|\mathbf{b}_n + \Delta \mathbf{x}_n\|^2$, where $\Delta \mathbf{x}_n \triangleq \Delta \mathbf{v}_1^H(n)\mathbf{\Upsilon}_i$, are independent from each other. Hence,

$$\Pr\{\|\epsilon_i\|^2 > \theta\} \geq \prod_{n=1}^{N} \xi_n, \tag{122}$$

where $\xi_n \triangleq \Pr\{\mu_n > \theta\}$. $\xi_n$ can be lower-bounded as follows:

$$\xi_n \overset{(a)}{\geq} 1 - \Pr\left\{\bigcup_{l=1}^{L_i} \|\mathbf{x}_n - \mathbf{d}_l\|^2 \leq \theta\right\}$$

$$\overset{(b)}{\geq} \max\left(0, 1 - \sum_{l=1}^{L_i} \Pr\left\{\|\mathbf{x}_n - \mathbf{d}_l\|^2 \leq \theta\right\}\right), \tag{123}$$

where $\mathbf{c}_l$, $l = 1, \cdots, L_i$, are the corresponding quantization code words for the quantization of $\mathbf{x}_n \triangleq \mathbf{v}_1^H(n)\mathbf{\Upsilon}_i$, and $\mathbf{d}_l \triangleq \mathbf{c}_l - \mathbf{b}_n$. $(a)$ comes from the fact that all the quantization bits are not necessarily utilized for the quantization of $\mathbf{x}_n$ [7], and $(b)$ results from the union bound for the probability.

Since the columns of $\mathbf{\Upsilon}_i$, namely $\{\mathbf{\Phi}_{\pi(j)}\}_{j=i+1}^M$, are semi-orthogonal to each other, $\mathbf{x}_n \triangleq \mathbf{v}_1^H(n)\mathbf{\Upsilon}_i$ can be approximated by $\mathbf{y}_n$, which denotes the projection of $\mathbf{v}_1(n)$ over the $(M - i)$-dimensional sub-space spanned by $\{\mathbf{\Phi}_{\pi(j)}\}_{j=i+1}^M$. More precisely,

$$\mathbf{x}_n \sim \mathbf{y}_n \left[\mathbf{I} + o(\mathbf{I})\right]. \tag{124}$$

As $\mathbf{v}_1(n)$ is an isotropically distributed unit vector in $\mathbb{C}^{1 \times M}$, the pdf of $\mathbf{y}_n$ can be computed from [28] as

$$p(\mathbf{y}_n) = \frac{(M-1)!}{\pi^{M-i}(i-1)!} \left(1 - \|\mathbf{y}_n\|^2\right)^{i-1}, \quad \|\mathbf{y}_n\| \leq 1. \tag{125}$$

---

[7] In fact, if we denote the original quantization code words, utilized for the quantization of $\mathbf{H}_n$, by $\{\mathbf{e}_l\}_{l=1}^{L_i}$, we can write $\mathbf{c}_l = f(\mathbf{e}_l)$, $1 \leq l \leq L_i$, where $f(.)$ is a mapping which depends on the quantization method. Since the mapping $f(.)$ is not necessarily one-to-one, it follows that the number of distinct elements in the set $\{\mathbf{c}_l\}_{l=1}^{L_i}$ is at most $L_i$.

October 9, 2018                                                                                                                                           DRAFT

Combining (124) and (125), $\Pr\{\|\mathbf{x}_n - \mathbf{d}_l\|^2 \leq \theta\}$ can be computed as

$$\begin{aligned}
\Pr\{\|\mathbf{x}_n - \mathbf{d}_l\|^2 \leq \theta\} &= \int_{C_{M-i}(\mathbf{d}_l,\sqrt{\theta})} p(\mathbf{x}_n)\mathrm{d}\mathbf{x}_n \\
&\stackrel{(124)}{\approx} \int_{C_{M-i}(\mathbf{d}_l,\sqrt{\theta})} p(\mathbf{y}_n)\mathrm{d}\mathbf{y}_n \\
&\stackrel{(a)}{\leq} \frac{(M-1)!}{\pi^{M-i}(i-1)!} \int_{C_{M-i}(\mathbf{d}_l,\sqrt{\theta})} \mathrm{d}\mathbf{y}_n \\
&= \frac{(M-1)!}{\pi^{M-i}(i-1)!} \mathrm{vol}\left(C_{M-i}(\mathbf{d}_l,\sqrt{\theta})\right) \\
&\stackrel{(b)}{=} \binom{M-1}{i-1}\theta^{M-i},
\end{aligned} \quad (126)$$

where $C_m(\mathbf{t},r)$ denotes the $m$-dimensional sphere (in the complex space) centered at $\mathbf{t}$ with radius $r$, and $\mathrm{vol}(\mathbf{v})$ denotes the volume of the region $\mathbf{v}$. $(a)$ comes from the fact that that from (125), $p(\mathbf{y}_n) \leq \frac{(M-1)!}{\pi^{M-i}(i-1)!}$, and $(b)$ results from the fact that the volume of a sphere with radios $d$ in the $m$-dimensional complex space is equal to $\frac{\pi^m}{m!}d^{2m}$. Substituting (126) in (123), we have

$$\xi_n \geq \max\left(0, 1 - L_i\binom{M-1}{i-1}\theta^{M-i}\right). \quad (127)$$

Substituting in (122), Lemma 4 easily follows.

∎

In Lemma 3, we have shown that in order to achieve the maximum sum-rate, in the case $K < M - i + 1$, we must have $\|\epsilon_i\|^2 \sim o\left(\frac{1}{P \ln N}\right)$ and in the case $K \geq M - i + 1$, we must have $\|\epsilon_i\|^2 \sim o(1)$, with probability one. In other words, in the first case,

$$\Pr\left\{\|\epsilon_i\|^2 > \frac{1}{P \ln N}\right\} \sim o(1), \quad (128)$$

and in the second case,

$$\Pr\left\{\|\epsilon_i\|^2 > 1\right\} \sim o(1). \quad (129)$$

Combining the above equations with (119), it follows that for the user with the encoding order i, such that $i \leq M - K$, we must have

$$\left(1 - L_i\binom{M-1}{i-1}\left[\frac{1}{P \ln N}\right]^{M-i}\right)^N \sim o(1) \Rightarrow L_i \sim \omega\left(\frac{[P \ln N]^{M-i}}{N}\right), \quad (130)$$

and for the users with the encoding order greater than $M - K$,

$$L_i \sim \omega\left(\frac{1}{N}\right). \quad (131)$$

Therefore, in the case of $K < M$, the total amount of feedback can be written as

$$\mathbb{E}\{\mathcal{F}_Q\} \stackrel{(a)}{\geq} \mathbb{E}\{\mathcal{N}_Q\} + \sum_{i=1}^{M-K} [\log_2(L_i)]^+$$

$$\stackrel{(b)}{\sim} \ln\ln(P \ln N) + \omega(1) + \sum_{i=1}^{M-K} \left[(M-i)\ln(P \ln N) - \ln N + \omega(1)\right]^+, \quad (132)$$

where $\mathcal{N}_Q$ denotes the number of users who send feedback to the BS. $(a)$ comes from the fact that at least $\mathcal{N}_Q$ users send one bit and $(M-K)$ users each send $[\log_2(L_i)]^+$ bits to the BS, where $L_i$ is computed from (130). $(b)$ results from (79) and (130).

In the case of $K = M$, (131) does not impose any constraints on $L_i$. Hence, the total amount of feedback can be lower-bounded as

$$\mathbb{E}\{\mathcal{F}_Q\} \geq \mathbb{E}\{\mathcal{N}_Q\}$$

$$\sim \ln\ln\ln N + \omega(1), \quad (133)$$

which completes the proof of Theorem 10.

∎

Although the above theorem gives us the necessary conditions for the amount of feedback to achieve the maximum sum-rate, the achievability of those conditions is not clear. A subsequent theorem gives the sufficient condition for achieving the maximum sum-rate.

From the above theorem the following observations can be made:

i) In the case of $K < M$, for the asymptotic scenario of $P \to \infty$, the minimum amount of feedback per user in order to achieve the maximum sum-rate grow logarithmically with SNR. This logarithmic growth is also shown for the fixed-size networks in [10], when the BS performs ZFBF. Moreover, for the fixed SNR scenario, this theorem implies that the minimum amount of feedback bits per user does not need to grow with $N$, which agrees with the result of Theorem 5, where we showed that the maximum sum-rate is achievable by a fixed amount of feedback per user.

ii) The more interesting observation is that, in the case of $K = M$, the above theorem does not impose any constraints on the minimum amount of feedback bits per user, even for the asymptotic scenario of $P \to \infty$. One may argue that this is not surprising as in this case, the transmitter can select the user which maximizes the single-user capacity (with a fixed amount of feedback per user, regardless of SNR), and communicating with that user, without knowing its channel. In [21], we have shown that this argument is not valid, as $\lim_{N \to \infty} \mathcal{R}_{\text{Opt}} - \mathcal{R}_{\text{TDMA}} = M \ln M$. In

3737373737

other words, there is a constant gap between the achieving sum-rate and the maximum sum-rate. In fact, the reason that this case differs form the previous case is the "interference hiding". Since each user has $M$ coordinates and the number of interfering coordinates is $M-1$, the transmitter can wisely hide the interference coordinates in the null-space of the signal coordinate, and thus the receiver does not see any interference. As a result, unlike the previous case, the total amount of feedback does not grow with SNR.

**Theorem 11** *The sufficient condition for achieving the maximum sum-rate, such that $\lim_{N,P\to\infty} \mathcal{R}_{\mathrm{Opt}} - \mathcal{R} = 0$, in the case of $K < M$ is*

$$\mathbb{E}\{\mathcal{F}_Q\} \sim [M(M-1)\ln P - M(K-1)\ln\ln N - o(\ln N)]^+ + \omega(\ln\ln(P\ln N)), \quad (134)$$

*and in the case of $K = M$ is*

$$\mathbb{E}\{\mathcal{F}_Q\} \sim M\ln\ln\ln N + \omega(1). \quad (135)$$

**Proof** - The proof is based on the two algorithms given in the following, in the cases $K < M$ and $K = M$. We show that by using these algorithms one can achieve the maximum sum-rate throughput of the system in each case, while the total amount of feedback satisfies (134) and (135), respectively.

Case $K < M$:

Consider the following algorithm:

1. Set the thresholds $t$, $\beta$, and $\epsilon$.
2. Define
$$\mathcal{S}_0 = \{k| \quad \lambda_{\max}(k) > t\},$$
   where $\lambda_{\max}(k)$ is the the maximum singular value of the $k$th user.
3. All users in $\mathcal{S}_0$ quantize the eigenvector corresponding to the maximum singular value of their channel matrix, denoted by $\mathbf{v}_k$, using the quantization code book $\mathcal{C} = \{\mathbf{c}_1, \cdots, \mathbf{c}_{2^B}\}$, where $\{\mathbf{c}_l\}_{l=1}^{2^B}$ are i.i.d. unit vectors with uniform distribution (RVQ). The quantized vector of $\mathbf{v}_k$, denoted by $\widehat{\mathbf{v}}_k$ is selected as
$$\widehat{\mathbf{v}}_k = \arg\max_{\mathbf{c}_l \in \mathcal{C}} |\mathbf{v}_k^H \mathbf{c}_l|.$$
4. All the users in the set
$$\mathcal{S}_1 = \left\{k \in \mathcal{S}_0 \middle| \quad |\mathbf{v}_k^H \widehat{\mathbf{v}}_k|^2 > 1 - \epsilon\right\}$$

send one bit to the BS. The BS selects one user in $\mathcal{S}_1$ at random and inform this user ($s_1$) to feed back its eigenvector. User $s_1$ feeds back the quantization index corresponding to its eigenvector to the BS. The BS sends this index to all the users in the set $\mathcal{S}_1 - \{s_1\}$.

5. For $m = 2$ to $M$ the following steps are repeated:
   - Define $\mathcal{S}_m = \left\{ k \in \mathcal{S}_{m-1} \middle| \ |\mathbf{v}_k^H \widehat{\mathbf{v}}_{s_{m-1}}|^2 < \beta \right\}$. All users in $\mathcal{S}_m$ send one bit to the BS.
   - The BS selects one user in $\mathcal{S}_m$ at random and informs this user ($s_m$) to feed back its corresponding eigenvector.
   - User $s_m$ feeds back the quantization index corresponding to its eigenvector to the BS. The BS sends this index to all the users in the set $\mathcal{S}_m - \{s_m\}$.

6. After selecting the users and receiving their quantized eigenvectors, the BS forms the beams $\{\mathbf{\Phi}_{s_m}\}_{m=1}^{M}$, such that $\mathbf{\Phi}_{s_m}$ is in the null-space of $\widehat{\mathbf{v}}_{s_j}$, $j \neq m$ (Zero-Forcing Beam-Forming). In other words, $\mathbf{\Phi}_{s_m}^H \widehat{\mathbf{v}}_{s_j} = 0$, $\forall j \neq m$.

7. The BS forms the transmitted signal as

$$\mathbf{x} = \sum_{j=1}^{M} \mathbf{\Phi}_{s_j} x_{s_j}, \tag{136}$$

where $x_{s_j} \sim \mathcal{CN}(0, \frac{P}{M})$ is the intended signal for the user $s_j$.

8. At the receiver $s_m$, the received vector $\mathbf{y}_{s_m}$ is multiplied by $\mathbf{u}_{s_m}^H$, where $\mathbf{u}_{s_m}$ denotes the left eigenvector corresponding to the maximum eigenvalue of the user $s_m$, to form $r_{s_m} = \mathbf{u}_{s_m}^H \mathbf{y}_{s_m}$. Then, the decoding is performed.

Defining the event $\mathcal{Q} \triangleq \bigcap_{m=1}^{M} \{|\mathcal{S}_m| \neq 0\}$, the sum-rate can be upper-bounded as

$$\begin{aligned} \mathcal{R} &= \Pr\{\mathcal{Q}\} \mathcal{R}_{\mathcal{Q}} + \Pr\{\mathcal{Q}^C\} \mathcal{R}_{\mathcal{Q}^C} \\ &\geq \Pr\{\mathcal{Q}\} \mathcal{R}_{\mathcal{Q}} \\ &\stackrel{(a)}{\geq} \left[1 - \sum_{m=1}^{M} \Pr\{|\mathcal{S}_m| = 0\}\right] \mathcal{R}_{\mathcal{Q}}, \end{aligned} \tag{137}$$

where $\mathcal{R}_{\mathcal{Q}}$ denotes the average sum-rate conditioned on $\mathcal{Q}$ and $(a)$ comes from the union bound for the probability. To compute $\mathcal{R}_{\mathcal{Q}}$, we calculate the rate of each user conditioned on $\mathcal{Q}$. For this purpose, the received signal by the $s_m$th user is simplified as follows:

$$
\begin{aligned}
r_{s_m} &= \mathbf{u}_{s_m}^H \mathbf{y}_{s_m} \\
&= \mathbf{u}_{s_m}^H [\mathbf{H}_{s_m} \mathbf{x} + \mathbf{n}_{s_m}] \\
&\stackrel{(a)}{=} \sqrt{\lambda_{\max}(s_m)} \mathbf{v}_{s_m}^H \mathbf{x} + z_{s_m} \\
&= \sqrt{\lambda_{\max}(s_m)} \mathbf{v}_{s_m}^H \mathbf{\Phi}_{s_m} x_{s_m} + \sum_{j \neq m} \sqrt{\lambda_{\max}(s_m)} \mathbf{v}_{s_m}^H \mathbf{\Phi}_{s_j} x_{s_j} + z_{s_m},
\end{aligned} \qquad (138)
$$

where $z_{s_m} \sim \mathcal{CN}(0,1)$ is AWGN and $(a)$ comes from writing SVD for $\mathbf{H}_{s_m}$. In the above equation, the first term contains the desired signal and the rest are the interference and noise terms. Hence, the rate of this user can be written as

$$
\mathcal{R}_{s_m} = \mathbb{E} \left\{ \ln \left( 1 + \frac{\frac{P}{M} \lambda_{\max}(s_m) \left| \mathbf{v}_{s_m}^H \mathbf{\Phi}_{s_m} \right|^2}{\sum_{j \neq m} \frac{P}{M} \lambda_{\max}(s_m) \left| \mathbf{v}_{s_m}^H \mathbf{\Phi}_{s_j} \right|^2 + 1} \right) \right\}. \qquad (139)
$$

We can write

$$
\mathbf{v}_{s_m} = \alpha_{s_m}^{\parallel} \widehat{\mathbf{v}}_{s_m} + \widehat{\mathbf{v}}_{s_m}^{\perp}, \qquad (140)
$$

where $\alpha_{s_m}^{\parallel} \triangleq \widehat{\mathbf{v}}_{s_m}^H \mathbf{v}_{s_m}$ and $\widehat{\mathbf{v}}_{s_m}^{\perp}$ is the projection of $\mathbf{v}_{s_m}$ over the sub-space perpendicular to $\widehat{\mathbf{v}}_{s_m}$. Using the above equation, $\left| \mathbf{v}_{s_m}^H \mathbf{\Phi}_{s_j} \right|^2$ can be written as

$$
\begin{aligned}
\left| \mathbf{v}_{s_m}^H \mathbf{\Phi}_{s_j} \right|^2 &= \left| \left( \alpha_{s_m}^{\parallel} \widehat{\mathbf{v}}_{s_m} + \widehat{\mathbf{v}}_{s_m}^{\perp} \right)^H \mathbf{\Phi}_{s_j} \right|^2 \\
&\stackrel{(a)}{=} \left| \left( \widehat{\mathbf{v}}_{s_m}^{\perp} \right)^H \mathbf{\Phi}_{s_j} \right|^2 \\
&\leq \left\| \widehat{\mathbf{v}}_{s_m}^{\perp} \right\|^2 \\
&= 1 - \left| \widehat{\mathbf{v}}_{s_m}^H \mathbf{v}_{s_m} \right|^2,
\end{aligned} \qquad (141)
$$

where $(a)$ comes from the fact that $\widehat{\mathbf{v}}_{s_m}^H \mathbf{\Phi}_{s_j} = 0$, $j \neq m$, by the algorithm. Conditioned on $\mathcal{Q}$, we have $\lambda_{\max}(s_m) > t$ and $\left| \widehat{\mathbf{v}}_{s_m}^H \mathbf{v}_{s_m} \right|^2 > 1 - \epsilon$. Therefore, the rate of the $s_m$th user, conditioned on $\mathcal{Q}$, can be lower-bounded as

$$
\mathcal{R}_{s_m | \mathcal{Q}} \geq \ln \left( 1 + \frac{\frac{Pt}{M} \left| \mathbf{v}_{s_m}^H \mathbf{\Phi}_{s_m} \right|^2}{1 + \frac{Pt \epsilon (M-1)}{M}} \right). \qquad (142)
$$

In the Appendix, we have shown that having $\beta \sim o(1)$ and $\epsilon \sim o(1)$ guarantees $\left| \mathbf{v}_{s_m}^H \mathbf{\Phi}_{s_m} \right|^2 \sim 1 + o(1)$. Having this, it follows that choosing $t \sim \ln N + o(\ln N)$ and $\epsilon \sim o\left(\frac{1}{P \ln N}\right)$ incurs $\mathcal{R}_{s_m | \mathcal{Q}} \sim \ln \left( 1 + \frac{P}{M} \ln N + o(\ln N) \right)$. Similarly, we can show that the same rate is achievable for the other selected users. Hence, $\mathcal{R}_{\mathcal{Q}} \sim M \ln \left( 1 + \frac{P}{M} \ln N + o(\ln N) \right)$ and as a result,

October 9, 2018DRAFT

$\lim_{P,N \to \infty} \mathcal{R}_{\text{Opt}} - \mathcal{R}_{\mathcal{Q}} = 0$. Using this fact and (137), it follows that the sufficient condition to achieve $\lim_{P,N \to \infty} \mathcal{R}_{\text{Opt}} - \mathcal{R} = 0$ is $\left[\sum_{m=1}^{M} \Pr\{|\mathcal{S}_m| = 0\}\right] \mathcal{R}_{\mathcal{Q}} \sim o(1)$, which incurs $\Pr\{|\mathcal{S}_m| = 0\} \sim o\left(\frac{1}{\ln(P \ln N)}\right)$. Since $\mathcal{S}_M \subseteq \mathcal{S}_{M-1} \subseteq \cdots \subseteq \mathcal{S}_1$, it suffices to consider only $\mathcal{S}_M$. Defining $q_k \triangleq \Pr\{k \in \mathcal{S}_M\}$ for a randomly chosen user $k$, we have

$$q_k = \Pr\left\{\lambda_{\max}(k) > t, |\mathbf{v}_k^H \widehat{\mathbf{v}}_{s_m}|^2 < \beta, m = 1, \cdots, M - 1, |\mathbf{v}_k^H \widehat{\mathbf{v}}_k|^2 > 1 - \epsilon\right\}. \quad (143)$$

Since the events $\mathcal{A}_1 \triangleq \{\lambda_{\max}(k) > t\}$, $\mathcal{A}_2 \triangleq \{|\mathbf{v}_k^H \widehat{\mathbf{v}}_{s_m}|^2 < \beta, m = 1, \cdots, M - 1\}$ and $\mathcal{A}_3 \triangleq \{|\mathbf{v}_k^H \widehat{\mathbf{v}}_k|^2 > 1 - \epsilon\}$ are independent of each other, $q_k$ can be written as $\prod_{i=1}^{3} q_{ki}$, where $q_{ki} \triangleq \Pr\{\mathcal{A}_i\}$. We have

$$\begin{aligned} q_{k1} &\overset{(a)}{\sim} \Theta\left(e^{-t} t^{M+K-2}\right), \\ q_{k2} &\overset{(b)}{\sim} \Theta(\beta^{M-1}), \end{aligned} \quad (144)$$

where $(a)$ comes from [20], and $(b)$ comes from [21]. Furthermore,

$$\begin{aligned} q_{k3} &= 1 - \Pr\left\{|\mathbf{v}_k^H \widehat{\mathbf{v}}_k|^2 < 1 - \epsilon\right\} \\ &= 1 - \prod_{l=1}^{L} \Pr\left\{|\mathbf{v}_k^H \mathbf{c}_l|^2 < 1 - \epsilon\right\} \\ &\overset{(a)}{=} 1 - \left(1 - \epsilon^{M-1}\right)^L \\ &\sim 1 - e^{-L\epsilon^{M-1}} \\ &\leq L\epsilon^{M-1}, \end{aligned} \quad (145)$$

where $L \triangleq 2^B$ and $(a)$ results from [21], Appendix C. Combining (144) and (145), we can write

$$\begin{aligned} \Pr\{|\mathcal{S}_M| = 0\} &\approx (1 - q_k)^N \\ &= (1 - q_{k1} q_{k2} q_{k3})^N \\ &\gtrsim \left[1 - \Theta\left(e^{-t} t^{M+K-2} \beta^{M-1} L \epsilon^{M-1}\right)\right]^N \\ &\sim \exp\left\{-\Theta\left(N e^{-t} t^{M+K-2} \beta^{M-1} L \epsilon^{M-1}\right)\right\}. \end{aligned} \quad (146)$$

Hence, in order to have $\Pr\{|\mathcal{S}_M| = 0\} \sim o\left(\frac{1}{\ln(P \ln N)}\right)$, it suffices to have

$$L \sim \Theta\left((\ln \ln(P \ln N) + \omega(1)) (\beta \epsilon)^{-(M-1)} N^{-1} e^t t^{-(M+K-2)}\right). \quad (147)$$

Choosing $\beta \sim o(1)$, $t = (1-\alpha)\ln N$, and $\epsilon = \frac{\delta}{P \ln N}$, where $\alpha, \delta \sim o(1)$, and substituting in the above equation, we obtain

$$\begin{aligned} L &\sim \Theta\Big( (\ln\ln(P\ln N) + \omega(1)) [P\ln N]^{M-1}(\beta\delta)^{-(M-1)} N^{-\alpha}[\ln N]^{-(M+K-2)} \Big) \\ &\sim \Theta\Big( (\ln\ln(P\ln N) + \omega(1)) P^{M-1}[\ln N]^{-(K-1)}(\beta\delta)^{-(M-1)} N^{-\alpha} \Big). \end{aligned} \quad (148)$$

Having $B = \lceil \log_2(L) \rceil^+$, yields

$$B \sim [(M-1)\ln P - (K-1)\ln\ln N + \ln\ln\ln(P\ln N) + \omega(1) - o(\ln N)]^+. \quad (149)$$

Using the above equation, the total amount of feedback can be written as

$$\begin{aligned} \mathbb{E}\{\mathcal{F}_Q\} &= MB + \sum_{m=1}^{M} \mathbb{E}\{|\mathcal{S}_m|\} \\ &= MB + \sum_{m=1}^{M} (N-m+1)\Pr\{k \in \mathcal{S}_m\} \\ &\overset{(a)}{\sim} MB + \omega(\ln\ln(P\ln N)) \\ &\sim [M(M-1)\ln P - M(K-1)\ln\ln N - o(\ln N)]^+ + \omega(\ln\ln(P\ln N)), \quad (150) \end{aligned}$$

where $(a)$ comes from the fact that selecting $L$ as in (147), results in $N\Pr\{k \in \mathcal{S}_M\} \sim \ln\ln(P\ln N) + \omega(1)$, and hence, $N\Pr\{k \in \mathcal{S}_m\} \sim N\Pr\{k \in \mathcal{S}_M\}\beta^{m-M} \sim \omega(\ln\ln(P\ln N))$.

Case $K = M$:

Consider the following algorithm:

1. Set the thresholds $t$ and $\epsilon$.

2. Define
$$\mathcal{S}_0 = \{k | \quad \lambda_{\max}(k) > t\},$$

where $\lambda_{\max}(k)$ is the the maximum singular value of the $k$th user.

3. The BS selects a unit vector $\mathbf{\Phi}_{s_1}$ at random and sends this vector to all users in $\mathcal{S}_0$.

4. All the users in the set
$$\mathcal{S}_1 = \left\{ k \in \mathcal{S}_0 \Big| \quad |\mathbf{v}_k^H \mathbf{\Phi}_{s_1}|^2 > 1 - \epsilon \right\},$$

where $\mathbf{v}_k$ denotes the eigenvector corresponding to the maximum eigenvalue of user $k$, send one bit to the BS. The BS selects one user in $\mathcal{S}_1$ at random indexed by $s_1$.

5. For $m = 2$ to $M$ the following steps are repeated:

- The BS selects a unit vector $\boldsymbol{\Phi}_{s_m}$ such that it is orthogonal to the previously chosen vectors $\{\boldsymbol{\Phi}_{s_j}\}_{j=1}^{m-1}$, and sends it to the users in $\mathcal{S}_0$.
- Define $\mathcal{S}_m = \left\{ k \in \mathcal{S}_0 \middle| \ |\mathbf{v}_k^H \boldsymbol{\Phi}_{s_m}|^2 > 1 - \epsilon \right\}$. All users in $\mathcal{S}_m$ send one bit to the BS.
- The BS selects one user in $\mathcal{S}_m$ at random indexed by $s_m$.

6. The BS forms the transmitted signal as

$$\mathbf{x} = \sum_{m=1}^{M} \boldsymbol{\Phi}_{s_m} x_{s_m}, \tag{151}$$

where $x_{s_m} \sim \mathcal{CN}(0, \frac{P}{M})$ is the intended signal for the user $s_m$.

7. At the receiver $s_m$, the received vector is multiplied by $\mathbf{R}_{s_m}^{-1/2}$, where

$$\mathbf{R}_{s_m} \triangleq \mathbf{I} + \sum_{j \neq m} \frac{P}{M} \mathbf{H}_{s_m} \boldsymbol{\Phi}_{s_j} \boldsymbol{\Phi}_{s_j}^H \mathbf{H}_{s_m}^H,$$

to form $\mathbf{r}_{s_m} = \mathbf{R}_{s_m}^{-1/2} \mathbf{y}_{s_m}$. Then, the decoding is performed.

As can be observed, this algorithm is very similar to the previous algorithm, with the difference in the quantization code book and decoding. In this algorithm, the quantization code book contains only one code word at each step, which is variable and decided by the BS, while in the previous algorithm the quantization code book is fixed and the number of code words grow with SNR. Moreover, the receiver uses all coordinates for decoding the signal, while in the previous algorithm the decoding is only performed in one coordinate. In fact, in the case of $K < M$, using all the coordinates does not provide any gain, while in the case of $K = M$, it does. In the case of $K = M$, if any of the sets $\mathcal{S}_m$, $m = 1, \cdots, M$, is empty, the BS selects any user at random and communicates with that user, setting the transmit covariance matrix equal to $\frac{P}{M}\mathbf{I}$. This provides a rate scaling as $M \ln P$, without requiring any amount of feedback.

Defining the event $\mathcal{Q} \triangleq \bigcap_{m=1}^{M} \{|\mathcal{S}_m| \neq 0\}$, similar to (137), we can write

$$\begin{aligned} \mathcal{R} &= \Pr\{\mathcal{Q}\}\mathcal{R}_\mathcal{Q} + [1 - \Pr\{\mathcal{Q}\}] \mathcal{R}_{\text{RS}}^{\mathcal{Q}^C} \\ &= \mathcal{R}_\mathcal{Q} - [1 - \Pr\{\mathcal{Q}\}] \left[ \mathcal{R}_\mathcal{Q} - \mathcal{R}_{\text{RS}}^{\mathcal{Q}^C} \right] \\ &\geq \mathcal{R}_\mathcal{Q} - \left( \sum_{m=1}^{M} \Pr\{|\mathcal{S}_m| = 0\} \right) \left[ \mathcal{R}_\mathcal{Q} - \mathcal{R}_{\text{RS}}^{\mathcal{Q}^C} \right], \end{aligned} \tag{152}$$

where $\mathcal{R}_{\text{RS}}^{\mathcal{Q}^C}$ denotes the achievable rate, when the BS selects one user at random and communicates with that user, conditioned on $\mathcal{Q}^C$. It is easy to show that $\mathcal{R}_{\text{RS}}^{\mathcal{Q}^C} \sim M \ln P + \Theta(1)$.

The rate of the user $s_m$, conditioned on $\mathcal{Q}$, can be computed as

$$\mathcal{R}_{s_m|\mathcal{Q}} = \mathbb{E}\left\{\ln\left|\mathbf{I} + \frac{P}{M}\mathbf{H}_{s_m}\boldsymbol{\Phi}_{s_m}\boldsymbol{\Phi}_{s_m}^H\mathbf{H}_{s_m}^H\mathbf{R}_{s_m}^{-1}\right|\,\Big|\,\mathcal{Q}\right\}. \tag{153}$$

For $\epsilon \sim o(1)$ and $t \sim \ln N$, and using the equations (110) and (117), it follows that

$$\mathcal{R}_{s_m|\mathcal{Q}} \gtrsim \mathbb{E}\left\{\ln\left(1 + \frac{P}{M}t(1-\epsilon)\mathbf{W}_{11}\right)\right\}$$
$$\sim \ln\left(1 + \frac{P}{M}\ln N[1+o(1)]\right), \tag{154}$$

where $\mathbf{W} = \mathbf{R}_{s_m}^{-1}$. Hence,

$$\mathcal{R}_\mathcal{Q} \sim M\ln\left(1 + \frac{P}{M}\ln N[1+o(1)]\right), \tag{155}$$

and as a result, $\mathcal{R}_{\text{Opt}} - \mathcal{R}_\mathcal{Q} \sim o(1)$. Therefore, having the fact that $\mathcal{R}_\mathcal{Q} - \mathcal{R}_{\text{RS}}^{\mathcal{Q}^C} \sim M\ln\ln N$, we can show that $\eta_m \triangleq \Pr\{|\mathcal{S}_m| \neq 0\} \sim o\left(\frac{1}{\ln\ln N}\right)$, $\forall m$, guarantees $\mathcal{R}_{\text{Opt}} - \mathcal{R} \sim o(1)$. $\eta_m$ can be written as $(1-q_m)^N$, where $q_m \triangleq \Pr\{k \in \mathcal{S}_m\}$, for a randomly chosen user $k$. $q_m$ can be computed as

$$q_m = \Pr\{\lambda_{\max}(k) > t\}\Pr\{|\mathbf{v}_k^H\boldsymbol{\Phi}_{s_m}|^2 > 1-\epsilon\}$$
$$\stackrel{(a)}{\sim} \frac{e^{-t}t^{M+K-2}}{\Gamma(M)\Gamma(K)}\epsilon^{M-1}, \tag{156}$$

where $(a)$ comes from [20] and [21]. Consequently,

$$\eta_m \sim \left[1 - \frac{e^{-t}t^{M+K-2}}{\Gamma(M)\Gamma(K)}\epsilon^{M-1}\right]^N$$
$$\sim e^{-N\frac{e^{-t}t^{M+K-2}}{\Gamma(M)\Gamma(K)}\epsilon^{M-1}}. \tag{157}$$

Choosing $\epsilon = \frac{1}{\ln N}$ and $t = \ln N + (K-1)\ln\ln N - \ln\ln\ln N - \ln\Gamma(M)\Gamma(K) - \omega\left(\frac{1}{\ln\ln N}\right)$ results in $\eta_m \sim o\left(\frac{1}{\ln\ln N}\right)$ and hence, having $\lim_{N,P\to\infty}\mathcal{R}_{\text{Opt}} - \mathcal{R} = 0$. The amount of feedback can be computed from

$$\mathbb{E}\{\mathcal{F}_Q\} = \mathbb{E}\left\{\sum_{m=1}^M |\mathcal{S}_m|\right\}$$
$$= N\sum_{m=1}^M q_m$$
$$\stackrel{(a)}{\approx} \sum_{m=1}^M \ln(\eta_m^{-1})$$
$$\sim M\ln\ln\ln N + \omega(1), \tag{158}$$

44where $(a)$ comes from the fact that $\eta_m = (1-q_m)^N \approx e^{-Nq_m}$.

∎

*Remark 1*- Comparing the necessary and sufficient conditions on the minimum amount of feedback for achieving the maximum sum-rate, it turns out that the proposed algorithm in the case of $K < M$ is asymptotically optimal by a constant multiplicative factor, in terms of the required amount of feedback, in the region $\ln P \sim \omega(\ln N)$. Moreover, in the case $K = M$, the proposed algorithm is optimal by a constant multiplicative factor, in terms of the required amount of feedback, for all ranges of SNR.

*Remark 2*- Comparing the two cases $K < M$ and $K = M$, it follows that the minimum amount of feedback in the first case grows logarithmically with SNR while in the second case it does not grow with SNR.

*Remark 3*- In the case of $K < M$, when $\ln P \sim \Omega(\ln N)$, it is possible to achieve the maximum sum-rate by using a finite-size quantization code book for all the users (Random Beam-Forming). However, in the case of $\ln P \sim \omega(\ln N)$, the size of the quantization code book must grow polynomially with SNR. In the case of $K = M$, it is possible to achieve the maximum sum-rate with finite rate quantization for all ranges of SNR. In other words, Random Beam-Forming is always optimal in this case. Note that, however, the decoding must be performed in all the coordinates.

*Remark 4*- The first algorithm can be considered as the generalization of Random Beam-Forming, when the number of beams vary with SNR. This algorithm is very similar to the algorithm proposed in [18], with the difference in limiting the number of candidate users and thus reducing the amount of feedback furthermore.

## IV. Conclusion

In this paper, the minimum required amount of feedback in order to achieve the maximum sum-rate capacity in a MIMO-BC with large number of users and different ranges of SNR is studied. In the fixed SNR and low SNR regimes, we have proved that to achieve the maximum sum-rate the total amount of feedback from the users to the BS must be infinity. Moreover, in the fixed SNR regime, in order to reduce the gap to the sum-rate capacity to zero, the amount of feedback must scale at least as $\ln \ln \ln N$, which is achievable by the Random Beam-Forming scheme introduced in [14]. Indeed, it is shown that the optimality of Random Beam-Forming scheme only holds for the region $\ln P \sim \Omega(\ln N)$. In the regime of $\ln P \sim \Omega(N)$, we consider two cases.

In the case of $K < M$, we prove that the minimum amount of feedback in order to reduce the gap between the achievable sum-rate and the maximum sum-rate to zero grows logaritmically with SNR, which is achievable by the "Generalized Random Beam-Forming" scheme proposed in [18]. In the case of $K = M$, we show that by using the Random Beam-Forming scheme with the total amount of feedback not growing with SNR, the maximum sum-rate capacity is achieved, provided that the decoding is performed in all the received coordinates.

## APPENDIX

To evaluate $\mathbf{v}_{s_m}^H \mathbf{\Phi}_{s_m}$, we define $\mathcal{P}_m$ as the sub-space defined by the vectors $\{\widehat{\mathbf{v}}_{s_i}\}_{i \neq m}$. We can write

$$\mathbf{v}_{s_m} = \mathbf{v}_{s_m}^{\|} + \mathbf{v}_{s_m}^{\perp}, \tag{159}$$

where $\mathbf{v}_{s_m}^{\|}$ is the projection of $\mathbf{v}_{s_m}$ over $\mathcal{P}_m$, and $\mathbf{v}_{s_m}^{\perp}$ is the projection of $\mathbf{v}_{s_m}$ over $\mathcal{P}_m^{\perp}$, and $\mathcal{P}_m^{\perp}$ denotes the sub-space perpendicular to $\mathcal{P}_m$. Since $\mathbf{\Phi}_m$ is perpendicular to all the vectors in the set $\{\widehat{\mathbf{v}}_{s_i}\}_{i \neq m}$, it belongs to $\mathcal{P}_m^{\perp}$, and we have

$$\begin{aligned}
\left|\mathbf{v}_{s_m}^H \mathbf{\Phi}_{s_m}\right|^2 &= \left|\left(\mathbf{v}_{s_m}^{\|} + \mathbf{v}_{s_m}^{\perp}\right)^H \mathbf{\Phi}_{s_m}\right|^2 \\
&= \left|\mathbf{\Phi}_{s_m}^H \mathbf{v}_{s_m}^{\perp}\right|^2 \\
&= \|\mathbf{v}_{s_m}^{\perp}\|^2 \\
&= 1 - \|\mathbf{v}_{s_m}^{\|}\|^2 \\
&\stackrel{(a)}{\gtrsim} 1 - \sum_{i \neq m} \left|\mathbf{v}_{s_m}^H \widehat{\mathbf{v}}_{s_i}\right|^2 \\
&\stackrel{(b)}{\geq} 1 - \sum_{i=1}^{m-1} \beta - \sum_{i=m+1}^{M} \left|\mathbf{v}_{s_m}^H \widehat{\mathbf{v}}_{s_i}\right|^2 \\
&\stackrel{(c)}{=} 1 - (m-1)\beta - \sum_{i=m+1}^{M} \left|\left(\alpha_m^{\|}\widehat{\mathbf{v}}_{s_m} + \widehat{\mathbf{v}}_{s_m}^{\perp}\right)^H \left(\gamma_i^{\|}\mathbf{v}_{s_i} + \mathbf{v}_{s_i}^{\perp}\right)\right|^2 \\
&\stackrel{(d)}{\geq} 1 - (m-1)\beta - \sum_{i=m+1}^{M} \left(\left|\widehat{\mathbf{v}}_{s_m}^H \mathbf{v}_{s_i}\right| + \|\widehat{\mathbf{v}}_{s_m}^{\perp}\| + \|\mathbf{v}_{s_i}^{\perp}\|\right)^2 \\
&\stackrel{(e)}{\geq} 1 - (m-1)\beta - \sum_{i=m+1}^{M} \left(\sqrt{\beta} + \sqrt{\mu_m} + \sqrt{\mu_i}\right)^2 \\
&\stackrel{(f)}{\geq} 1 - (m-1)\beta - 3\sum_{i=m+1}^{M} (\beta + \mu_m + \mu_i) \\
&\stackrel{(g)}{\geq} 1 - (3M - 2m - 1)\beta - 6(M - m)\epsilon. \tag{160}
\end{aligned}$$

In the above equation, $(a)$ follows from the fact that $\{\widehat{\mathbf{v}}_{s_i}\}_{i \neq m}$ form an semi-orthogonal basis for $\mathcal{P}_i$. To see this, we evaluate $\left|\widehat{\mathbf{v}}_{s_i}^H \widehat{\mathbf{v}}_{s_j}\right|^2$, $i, j \neq m$, for $i > j$. For this purpose, we write $\widehat{\mathbf{v}}_{s_i}$ as $\gamma_i^{\|}\mathbf{v}_{s_i} + \mathbf{v}_{s_i}^{\perp}$, in which $\mathbf{v}_{s_i}^{\perp}$ denotes the projection of $\widehat{\mathbf{v}}_{s_i}$ over the subspace perpendicular to $\mathbf{v}_{s_i}$, and $\gamma_i^\parallel \triangleq \mathbf{v}_{s_i}^H \widehat{\mathbf{v}}_{s_i}$. Then, we have

$$\begin{aligned}
\left|\widehat{\mathbf{v}}_{s_i}^H \widehat{\mathbf{v}}_{s_j}\right|^2 &= \left|\left(\gamma_i^\parallel \mathbf{v}_{s_i} + \mathbf{v}_{s_i}^\perp\right)^H \widehat{\mathbf{v}}_{s_j}\right|^2 \\
&\leq \left(\left|\gamma_i^\parallel\right| \left|\mathbf{v}_{s_i}^H \widehat{\mathbf{v}}_{s_j}\right| + \left|\left[\mathbf{v}_{s_i}^\perp\right]^H \widehat{\mathbf{v}}_{s_j}\right|\right)^2 \\
&\leq \left(\sqrt{\beta} + \|\mathbf{v}_{s_i}^\perp\|\right)^2 \\
&\leq \left(\sqrt{\beta} + \sqrt{\epsilon}\right)^2 \\
&\sim o(1),
\end{aligned} \qquad (161)$$

where the first inequality results from the fact that $|a + b|^2 \leq (|a| + |b|)^2$, $\forall a, b$, the second inequality follows from the facts that $\left|\gamma_i^\parallel\right| \leq 1$, $\left|\mathbf{v}_{s_i}^H \widehat{\mathbf{v}}_{s_j}\right| < \sqrt{\beta}$ (by the algorithm), and $\left|\left[\mathbf{v}_{s_i}^\perp\right]^H \widehat{\mathbf{v}}_{s_j}\right| \leq \|\mathbf{v}_{s_i}^\perp\|$, the third inequality results from the fact that $\|\mathbf{v}_{s_i}^\perp\|^2 = 1 - \left|\mathbf{v}_{s_i}^H \widehat{\mathbf{v}}_{s_i}\right|^2$, which is by the algorithm upper-bounded by $\epsilon$, and finally, the last line follows from the assumptions of $\epsilon \sim o(1)$ and $\beta \sim o(1)$.

The inequality $(b)$ in (160) comes from the fact that $\left|\mathbf{v}_{s_m}^H \widehat{\mathbf{v}}_{s_i}\right|^2 < \beta$ for $i < m$ by the algorithm. The equality $(c)$ results from writing $\mathbf{v}_{s_m}$ as $\alpha_m^\parallel \widehat{\mathbf{v}}_{s_m} + \widehat{\mathbf{v}}_{s_m}^\perp$ and $\widehat{\mathbf{v}}_{s_i}$ as $\gamma_i^\parallel \mathbf{v}_{s_i} + \mathbf{v}_{s_i}^\perp$ with the assumption of $\widehat{\mathbf{v}}_{s_m}^H \widehat{\mathbf{v}}_{s_m}^\perp = 0$, and $\mathbf{v}_{s_i}^H \mathbf{v}_{s_i}^\perp = 0$. Hence, it follows that $\alpha_m^\parallel = \widehat{\mathbf{v}}_{s_m}^H \mathbf{v}_{s_m}$, $\gamma_i^\parallel = \mathbf{v}_{s_i}^H \widehat{\mathbf{v}}_{s_i}$, $\|\widehat{\mathbf{v}}_{s_m}^\perp\|^2 = 1 - \left|\alpha_m^\parallel\right|^2$, and $\|\mathbf{v}_{s_i}^\perp\|^2 = 1 - \left|\gamma_i^\parallel\right|^2$. Inequality $(d)$ follows from the fact that $\left|\gamma_i^\parallel\right| < 1$, $\left|\alpha_m^\parallel\right| < 1$, $\left|\widehat{\mathbf{v}}_{s_m}^H \mathbf{v}_{s_i}^\perp\right| < \|\mathbf{v}_{s_i}^\perp\|$ and $\left|\mathbf{v}_{s_i}^H \widehat{\mathbf{v}}_{s_m}^\perp\right| < \|\widehat{\mathbf{v}}_{s_m}^\perp\|$. Inequality $(e)$ comes from the fact that $\left|\widehat{\mathbf{v}}_{s_m}^H \mathbf{v}_{s_i}\right|^2 < \beta$ for $i > m$ by the algorithm, and defining $\mu_m \triangleq \|\widehat{\mathbf{v}}_{s_m}^\perp\|^2 = 1 - \left|\mathbf{v}_{s_m}^H \widehat{\mathbf{v}}_{s_m}\right|^2$ and $\mu_i \triangleq \|\mathbf{v}_{s_i}^\perp\|^2 = 1 - \left|\mathbf{v}_{s_i}^H \widehat{\mathbf{v}}_{s_i}\right|^2$. Inequality $(f)$ comes from the fact that $\forall a, b, c$, $(a+b+c)^2 \leq 3(a^2+b^2+c^2)$, and finally, $(g)$ results from the fact that $\left|\mathbf{v}_{s_m}^H \widehat{\mathbf{v}}_{s_m}\right|^2 > 1-\epsilon$ for all $1 \leq m \leq M$. From the above equation, it can be observed that having $\beta \sim o(1)$ and $\epsilon \sim o(1)$ yields $\left|\mathbf{v}_{s_m}^H \mathbf{\Phi}_{s_m}\right|^2 \sim 1 - o(1)$.